\documentclass[1p,sort&compress,times]{elsarticle}
\usepackage{amsmath,amssymb}
\usepackage{graphicx}
\usepackage[breaklinks,colorlinks=true]{hyperref}

\usepackage{braket}

\newcommand{\rme}{\mathrm{e}}
\newcommand{\rmi}{\mathrm{i}}
\newcommand{\rmd}{\mathrm{d}}

\begin{document}

\title{Lefschetz-thimble inspired analysis of the
  Dykhne-Davis-Pechukas method and an application for the
  Schwinger Mechanism}

\author[tokyo,ipi]{Kenji Fukushima}
\ead{fuku@nt.phys.s.u-tokyo.ac.jp}
\author[tokyo]{Takuya Shimazaki}
\ead{shimazaki@nt.phys.s.u-tokyo.ac.jp}

\address[tokyo]{Department of Physics, The University of Tokyo, %
  7-3-1 Hongo, Bunkyo-ku, Tokyo 113-0033, Japan}
\address[ipi]{Institute for Physics of Intelligence (IPI), The University of Tokyo, %
  7-3-1 Hongo, Bunkyo-ku, Tokyo 113-0033, Japan}

\begin{abstract}
Dykhne-Davis-Pechukas (DDP) method is a common approximation
scheme for the transition probability in two-level quantum systems,
as realized in the Landau-Zener effect, leading to an exponentially
damping form comparable to the Schwinger pair production rate.  We
analyze the foundation of the DDP method using a modern complex
technique inspired by the Lefschetz-thimble method.  We derive an
alternative and more adaptive formula that is useful even when the DDP
method is inapplicable.  As a benchmark, we study the modified
Landau-Zener model and compare results from the DDP and our methods.
We then revisit a derivation of the Schwinger Mechanism of particle
production under electric fields using the DDP and our methods.  We find
that the DDP method gets worse for the Sauter type of short-lived
electric pulse, while our method is still a reasonable approximation.
We also study the Dynamically Assisted Schwinger Mechanism in two
methods.
\end{abstract}


\maketitle

\section{Introduction}

Nonadiabatic transitions are ubiquitous phenomena in physics problems
such as dielectric breakdown of a Mott
insulator~\cite{oka2010dielectric,oka2012nonlinear}, shortcuts to
adiabaticity in quantum
manipulation~\cite{berry2009transitionless,sinitsyn2018integrable},
population transfer in
optics~\cite{vasilev2009optimum,guerin2011optimal,zhang2019fast,ma2019high},
and so on.  In many-body systems theoretical treatments often reduce to
one-quasi-particle problem on energy levels formed by surrounding mean
fields as exemplified in the band theory of solids, the shell model of
nuclei, etc.  Then, the essence of quantum tunneling phenomena is in
effect modeled in a form of relatively simple system in Quantum
Mechanics.  In many-body systems of Quantum Field Theories
nonadiabatic transitions from an anti-particle (negative energy) state
to a particle (positive energy) state can be interpreted as pair particle
production, which is actually what we will address in details in this paper.
Several examples are found in the Schwinger
Mechanism~\cite{Sauter:1931zz,Heisenberg:1935qt,Schwinger:1951nm} (see Refs.~\cite{Dunne:2004nc,Gelis:2015kya} for reviews),
i.e., pair production under external electric field, and also in
high-energy nuclear collision where particle production occurs from strong
chromo-electromagnetic fields.  As long as spatial inhomogeneity is
assumed for such problems, we can compute the transition rate for each
momentum mode in the same way as two-level systems in Quantum
Mechanics.  Similarly, the Hawking radiation can be also treated as a
simple Quantum Mechanical problem~\cite{Parikh:1999mf}.  It is thus an
important theoretical challenge, not closed to Quantum Mechanics but
relevant generally to Quantum Field Theories, to consider handy and
multi-purpose formulas for estimating quantum transition amplitudes.

To attack this challenge there are various approaches, and here, let
us pick up one specific method called the Dykhne-Davis Pechukas (DDP)
method, which is of frequent choice particularly in condensed matter
physics fields (see,
Refs.~\cite{oka2010dielectric,oka2012nonlinear} for example).
In the early 1930's pioneering works paved a way for real-time quantum
dynamics beyond the adiabatic theorem.  Among them,
Landau and Zener especially derived an analytical formula for the transition
probability in a simple two-level quantum model which is known as the
Landau-Zener model today~\cite{landau1932theorie,zener1932non}.  Dykhne proposed
an approximation method with complexified time in such a way adaptive
for more general two-level quantum systems~\cite{dykhne1962adiabatic}.  Davis and
Pechukas derived the method in a clearer manner removing ad hoc
assumptions in Dykhne's formulation~\cite{davis1976nonadiabatic}, so the DDP
method aka the Landau-Dykhne method was named after these founders.

There are many preceding literatures on the DDP method;  examples
include the adiabatic limit~\cite{suominen1991adiabatic}, nonlinear
level-crossing models~\cite{vitanov1999nonlinear}, nonadiabatic
transitions in multi-level systems~\cite{wilkinson2000nonadiabatic},
and so on.  We note that more references are easily found for what is
called the Landau-Zener formula, but the DDP method is a more
sophisticated formulation than the Landau-Zener formula.  The most
non-trivial is the fact that the DDP method mystically reproduces the
exact analytical result for the Landau-Zener model, although the
derivation of the DDP method involves several approximating steps as
we will explain later.

Apart from technical details, at this point, we would remark that the
validity of the DDP method requires the following assumptions:
(1) The difference of two-level adiabatic
energies $\delta E(t)$ has a closing point $t_{c}\not\in\mathbb{R}$ in
complex-$t$ plane where $\delta E(t_c) = 0$ is satisfied.
(2) The $t$-contour can be deformed from the real axis to the DDP
choice (which will be specified later) without hitting any poles in
complex-$t$ plane.
When these assumptions hold, it is empirically known that the DDP
method is a good approximation, while there are a few exceptions due
to pathological behavior of
$\delta E(t\sim\pm\infty)$~\cite{laine1996adiabatic,vitanov1996non}.
In the present work we will pay our special attention on a model which
violates the condition (2) as mentioned above.  As soon as poles
hinder the deformation of the integration contour, the DDP method
breaks down and we must employ a more firm method taking proper
account of the complex structures.

The aim of the present work is to establish another approximate method
based on the Lefschetz-thimble method.  The Lefschetz-thimble method
is a modern complex analysis treating an integral with complexified
valuables.  The original targets of the Lefschetz-thimble method were
path integrals in quantum mechanics~\cite{Witten:2010zr}, quantum
gravity~\cite{witten2011analytic}, and quantum field theory.  Some
examples for recent applications are found for the QCD sign problem at
finite density~\cite{Cristoforetti:2012su,Cristoforetti:2013wha,Fujii:2013sra}
(which is investigated also in a more idealized one-site fermion
model~\cite{tanizaki2016lefschetz}), the real-time Feynman
path integral~\cite{tanizaki2014real,Cherman:2014sba}, the false
vacuum decay~\cite{andreassen2017precision}, and so on.
This method provides us with a clear prescription for complex saddle
points and attached contours as well as interesting physics
interpretations (see Ref.~\cite{Behtash:2015loa} for a review).  The
difference from standard saddle-point approximation is that, when
multiple saddle points appear, this method clearly tells us which
points we should take and which we should not.

In this work we develop the Lefschetz-thimble inspired method to
calculate the transition probability in two-level quantum systems.
Our approach based on the Lefschetz-thimble method has advantages over
the conventional DDP method.  First of all, our method is an
approximation under theoretical control and can be improved
systematically.  Sometimes the DDP method works artificially better
beyond the approximation limit as we discuss later.  It is, however,
hard to judge how reliable the DDP method is, \textit{a priori}, in
general.  Second, our method is not prevented by poles that obstruct
the DDP method.  Therefore, the applicability of our method is much
wider than the DDP method.  We explicitly demonstrate the above
advantages using the modified Landau-Zener model as a warm-up
exercise.

Armed with experiences of such concrete examples, we shall proceed to
another related problem of the Schwinger Mechanism. Although the
Schwinger Mechanism at finite temperature is an active research field
lately~\cite{Gould:2017fve,Draper:2018lyw,Gould:2018efv,Gould:2018ovk},
we limit ourselves to the exponent only of the production
rate at zero temperature, and will not calculate the prefactor
in this paper, for simplicity.  We can show that the DDP
method immediately leads to the famous Schwinger exponential factor
correctly for constant electric field.  We next consider a little more
non-trivial case with pulsed electric field in a Sauter-type form.  We
see that the complex structures are a little more complicated, but for
a pulse whose scale is smaller than the mass gap, we can still utilize
the DDP method to obtain the correct estimate.  Interestingly, for a
sharp pulse, on the other hand, the estimate from the DDP method gets
worse.  Finally, we will consider a combination of a pulse on top of
constant electric field.  In this case the exponent is significantly
suppressed for particular parameter regions, and this suppression
mechanism is commonly referred to as the Dynamically Assisted
Schwinger Mechanism~\cite{Schutzhold:2008pz}. The Dynamically Assisted
schwinger Mechanism helps to realize experimentally the Schwinger
Mechanism which has not been verified yet, so it has been actively
attracting attention~\cite{Dunne:2009gi,Orthaber:2011cm,Fey:2011if,Li:2014psw,Copinger:2016llk,Schneider:2016vrl,Torgrimsson:2017pzs,Torgrimsson:2017cyb}. Again,
the DDP method can reproduce the exponent
correctly as long as the pulse is broad enough.  We will make a
quantitative comparison between results from the fully numerical
calculation, the DDP method, and our method.

Throughout this work we adopt physical units of $\hbar=c=1$.

\section{Transition amplitude in two-level quantum systems}

We consider two-level systems as the simplest quantum mechanical
example.  Without loss of generality, we can reduce two-level problems
into the ones described by the following
Hamiltonian~\cite{berry1990histories},
\begin{equation}
  \label{eq:hamiltonian}
  H(t) = \frac{1}{2}\left[
  \begin{array}{cc}
    \alpha(t) & V(t)\\
    V(t) & -\alpha(t)
  \end{array}
  \right]\,,
\end{equation}
where $\alpha(t)$ and $V(t)$ are real-valued functions.  We define
adiabatic eigenvalues $E_\pm(t)$ and adiabatic eigenstates
$\ket{\chi_\pm(t)}$ obtained from $H(t)$ as
\begin{equation}
  H(t)\ket{\chi_\pm(t)} = E_\pm(t)\ket{\chi_\pm(t)}\,.
\end{equation}
We choose $E_+(t)\geq E_-(t)$ for $t\in\mathbb{R}$.  We denote a
physical state by $\ket{\psi(t)}$ that obeys the Schr\"{o}dinger
equation and expand $\ket{\psi(t)}$ in terms of $\ket{\chi_\pm(t)}$,
i.e.,
\begin{equation}
  \ket{\psi(t)} = \sum_{i=\pm} a_i(t)\, \rme^{-\rmi\mathcal{E}_i(t)}\ket{\chi_i(t)}
  \qquad \text{with} \qquad
  \mathcal{E}_\pm(t) = \int^t_0 \rmd t' E_\pm(t')\,.
\end{equation}
We derive the equations of motion for $a_\pm(t)$ from the
Schr\"{o}dinger equation as
\begin{equation}
  \label{eq:apluseq}
  \dot{a}_\pm(t) = \pm\eta(t) \, \rme^{\pm \rmi\Delta(t)}a_\mp(t)\,,
\end{equation}
where
\begin{align}
  \Delta(t) &= \mathcal{E}_+(t)-\mathcal{E}_-(t) = \int^t_0 \rmd t' \delta E(t')\,, \\
\ \eta(t) &= \braket{\chi_-(t)|\frac{\rmd}{\rmd t}|\chi_+(t)} = \frac{1}{2}\frac{V\dot{\alpha}-\dot{V}\alpha}{\alpha^2 + V^2}
\end{align}
with $\delta E(t)= E_{+}(t)-E_{-}(t)=\sqrt{\alpha(t)^2 + V(t)^2}$.
If we solve the time evolution
with initial conditions, $a_-(-\infty) = 1\,$ and $a_+(-\infty)=0$,
then $P=|a_+(\infty)|^2$ gives the transition probability from
$\ket{\chi_{-}(-\infty)}$ to $\ket{\chi_{+}(\infty)}$.

To this end of obtaining $P$, we integrate Eq.~\eqref{eq:apluseq} with
respect to $t$ to find,
\begin{equation}
  a_+(\infty)
  = \int_{-\infty}^\infty \rmd t\, a_-(t) \, \eta(t) \, \rme^{\rmi \Delta(t)}
  \simeq \int_{-\infty}^\infty \rmd t\, \rme^{F(t)} \,,
  \label{eq:aplus}
\end{equation}
where we defined $F(t) = \rmi\Delta(t) + \ln \eta(t)$.  To go to the
last expression we employed the first order truncation in the
iterative approximation, that is, we set $a_-(t) = 1$.  Physically
speaking, the time evolution should generally involve many alternate
transitions between $\ket{\chi_{\pm}(t)}$, but the above first order
truncation corresponds to an approximation by one time transition from
$\ket{\chi_{-}(t)}$ to $\ket{\chi_{+}(t)}$.  In other words this is
the \textit{definition} of quantity of our current interest, as is
also relevant to discussions on the Schwinger Mechanism (see the
difference between $f$ and $w$ in Sec.~IIB of
Ref.~\cite{Cohen:2008wz}).  At this point we make clear that
Eq.~\eqref{eq:aplus} is the starting point for further approximations
including the DDP method.

\subsection{DDP method}

\begin{figure}
    \centering
    \includegraphics[width=0.45\textwidth]{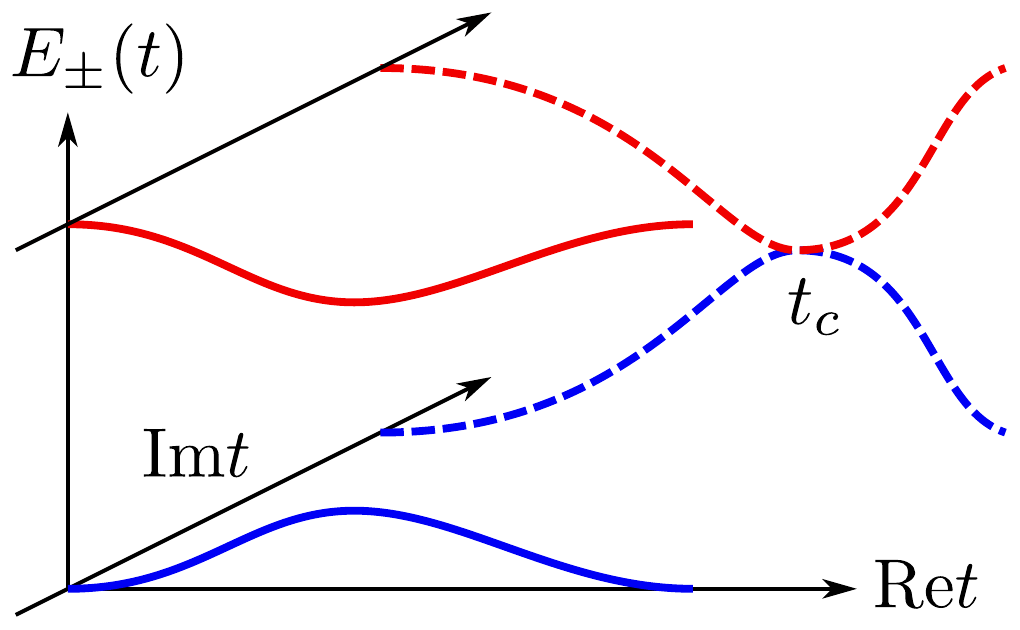}
    \caption{Schematic illustration of the closing point $t_c$ in the
      complex-$t$ plane.}
    \label{fig:closing}
\end{figure}

The DDP method is derived from Eq.~\eqref{eq:aplus} with deformed
integration contour in terms of complexified
$t$~\cite{davis1976nonadiabatic}.  In what follows below, we consider
the situation where there exists only one closing point, $t_c$, such
that $\delta E(t_{c})=0$ in upper complex-$t$ plane (see
Fig.~\ref{fig:closing} for a schematic illustration).
[Let us comment on generalization to multiple closing points in upper
complex-$t$ plane.  In such a case the DDP method prescribes that the
nearest closing point(s) contributes to the amplitude with proper
weights which are calculable.  Our proposed approach, explicated
shortly, is also applicable to the case with multiple closing points
in a similar fashion.]  Equivalently, this condition is translated into
\begin{equation}
  \frac{\partial \Delta}{\partial t} \biggr|_{t=t_c} = 0\,.
  \label{eq:t_c}
\end{equation}
Here, we used a partial derivative, for in later discussions we will
introduce spatial coordinates.

The second assumption in the DDP method, i.e., deformability of the
integration contour, needs some more explanations.  The complex-$t$
contour adopted in the DDP method is the one determined by
\begin{equation}
  {\rm Im}\Delta(t) = {\rm Im}\Delta(t_c)\,.
  \label{eq:contourDDP}
\end{equation}
The second assumption holds if there are no poles between the real
axis and this contour.  Still, we would note that the deformability is
subtle because the contour at infinity may have a non-vanishing
contribution.  Once these conditions for the deformability are
premised, the transition amplitude from $\ket{\chi_{-}(-\infty)}$ to
$\ket{\chi_{+}(\infty)}$ is approximated by
$a_{+}(\infty) \simeq \rme^{\rmi \Delta(t_c)}$ so that the transition
probability is
\begin{equation}
  P \simeq \rme^{-2{\rm Im}\Delta(t_c)} \,.
  \label{eq:DDP}
\end{equation}
With the above form, the DDP choice~\eqref{eq:contourDDP} could be
intuitive to some extent; only the imaginary part contributes to the
exponent, and the real part represents an oscillation phase.  Along
the contour of Eq.~\eqref{eq:contourDDP}, the stationary point
approximation at $t=t_c$ (if it corresponds to the stationary point,
which is not always the case, though) finds a dominant exponential
term that is Eq.~\eqref{eq:DDP}.  As we will closely discuss later,
Eq.~\eqref{eq:DDP} is a powerful tool to investigate general
transitional problems including quantum field theoretical problems
such as the Schwinger Mechanism in electric fields in a particular
regime.

The DDP method provides us with a fascinating interpretation for the
transitional processes.  Two adiabatic eigenenergies are typically gapped
as illustrated in Fig.~\ref{fig:closing}, and the transition needs quantum
tunneling.  Once the notion of time is extended to the complex plane,
however, it is possible to take a \textit{detour} to reach $t_c$ where
two states touch each other, as depicted in Fig.~\ref{fig:closing},
and thus the transition can occur classically there.  In some sense
the idea is analogous to the instanton calculus based on classical
trajectory with Euclidean time, but the current formulation is more
general in the while complex-$t$ plane.

\subsection{Complex analysis}

The DDP method is certainly interesting and useful, but unfortunately,
it may fail when it fails as we will see.  Thus, we shall consider an
independent approximation scheme here.  To approximate the
integral~\eqref{eq:aplus}, instead of the DDP method, we employ the
complex analysis inspired by recent developments of the
Lefschetz-thimble method.  One can say that the Lefschetz-thimble
method is a modern refinement of the steepest descent method.  We
assume that saddle points, $t_s$, of $F(t)$ are nondegenerate, i.e.,
\begin{equation}
  \frac{\partial F}{\partial t}\Biggr|_{t = t_s}
  = \Biggl[ \rmi \frac{\partial \Delta}{\partial t} + \eta(t)^{-1}
  \frac{\partial \eta}{\partial t} \Biggr]_{t=t_s}
   = 0\,,
  \qquad
  \frac{\partial^2 F}{\partial t^2}\Biggr|_{t = t_s} = |F''(t_{s})|\,\rme^{\rmi\phi} \neq 0\,,
\end{equation}
where $0\leq\phi<2\pi$.  We note that the difference between $t_c$ and $t_s$
results from the second term in the square parentheses involving $\eta$
[see Eq.~\eqref{eq:t_c}].
  We expand $F(t)$ around the saddle point,
$t_{s}$, as
\begin{equation}
  \label{eq:taylor}
  F(t) = F(t_{s}) + \frac{|F''(t_{s})|}{2}
  r^2 \rme^{\rmi(\phi + 2\theta)} + O(r^{3})\,.
\end{equation}
We note that we introduced the polar form, $t-t_{s} = r\rme^{\rmi \theta}$.
The contour along $\phi+2\theta = \pi$ around $t_{s}$ is directed
toward a steepest descent.  According to the Lefschetz-thimble method,
the original integration contour is decomposed into steepest descents
from saddle points.  The $i$-th saddle point, $t_{s,i}$, is attached
to a steepest descent $\mathcal{J}_i$ and a steepest ascent
$\mathcal{K}_i$.  We note that ${\rm Im}F(t) = \text{(const.)}$ holds
on $\mathcal{J}_i$, which makes a sharp contrast to the previous
condition~\eqref{eq:contourDDP} that keeps, apart from $\ln \eta(t)$,
${\rm Re}F(t)$ constant.
Now, it is a crucial process to specify which saddle
points should be taken into account to recover the original integral
\textit{exactly}.  It is known that $\{\mathcal{J}_{i}\} $ generate the 1st
relative homology $H_1(X,X_{-\infty},\mathbb{Z})$ when $X$ denotes
complex-$t$ plane and $X_T$ denotes
$\{t\in X| \mbox{Re}F(t) < T\}$~\cite{witten2011analytic}.
The bases of $H_1(X,X_{-\infty},\mathbb{Z})$ are called the Lefschetz
thimbles in the Picard-Lefschetz theory, which are nothing but
steepest descents in this case.  The original integration contour is
the real axis denoted here by $C_0$.  Then, the original integral is
recovered exactly if the deformed contour is chosen to be
\begin{equation}
  \label{eq:contoursum}
  C = \sum_i n_i \mathcal{J}_i\,,
\end{equation}
where the Morse index, $n_i\in\mathbb{Z}$, must be carefully determined,
and the calculational prescription is well established as
\begin{equation}
  n_i = \langle C_0, \mathcal{K}_i \rangle\,,
\end{equation}
where $\langle C_{0},\mathcal{K}_{i} \rangle$ counts the intersection
number between $C_{0}$ and $\mathcal{K}_{i}$ with orientation in
accord with
$\langle \mathcal{J}_i,\mathcal{K}_j\rangle = \delta_{ij}$.
Here, we emphasize that Eq.~\eqref{eq:contoursum} is the most
essential difference from the standard complex analysis which simply
prescribes contour deformations based on Cauchy's integral formula.
One might think that complex poles would make things complicated as is
the case for the DDP formula, but the appropriate choice of
$\mathcal{J}_i$ and $n_i$ already takes care of such complications.

With Eqs.~\eqref{eq:aplus}, \eqref{eq:taylor}, and
\eqref{eq:contoursum}, the saddle-point approximation leads to
\begin{equation}
  \label{eq:Lef}
  a_+(\infty) \simeq \sum_i n_i\, \rme^{\rmi\theta_i + F(t_{s,i})}\sqrt{\frac{2\pi}{|F''(t_{s,i})|}}
\end{equation}
after our performing the Gaussian integration with respect to $r$ on each
$\mathcal{J}_i$.  Here, again, we stress that the determination of $n_i$ is
crucially important in above formula~\eqref{eq:Lef}.
We will apply Eq.~\eqref{eq:Lef} to a two-level
quantum system in order to demonstrate how this formula works more
robustly than the DDP method in wider parameter regions.

\subsection{Benchmark test in the simplest example}

We note that the DDP method happens to reproduce the analytical
formula exactly in a special model called the Landau-Zener model.  The
diabatic energies are, however, divergent unphysically as
$t\to\pm\infty$ and the interaction between diabatic states is
constant in the Landau-Zener model.  It is thus desirable to augment
the Landau-Zener model with $t$-dependence so that the asymptotic
behavior becomes more treatable.  We consider the following
Hamiltonian,
\begin{equation}
  \label{eq:mLZ}
  H(t) = \frac{\Lambda}{2\tau\sqrt{1 + (t/T)^2}}\left(
  \begin{array}{cc}
  t/\tau & 1 \\
  1 & -t/\tau
  \end{array}\right)\,,
\end{equation}
where $\Lambda,\tau$, and $T$ are positive parameters.  In the limit
of $T\to\infty$, this model reduces to the Landau-Zener
model.  So, we call the above
model defined by Eq.~\eqref{eq:mLZ} the modified Landau-Zener model,
which is sometimes called the simple Lorentzian singularity
model~\cite{suominen1991adiabatic}.  It is easy to find the difference
of adiabatic energies given by
\begin{equation}
  \delta E(t) = \frac{T\Lambda}{\tau^2}\sqrt{\frac{t^2 +\tau^2}{t^2 + T^2}}\,.
\end{equation}
This expression of $\delta E(t)$ has a closing point at $t=i\tau$ and
a pole at $t=iT$ in upper complex-$t$ plane.  For $T>\tau$ the DDP
method is applicable since we can deform the integration contour
without hitting the pole.  In contrast, the DDP method breaks down for
$T\leq\tau$, while our method based on Eq.~\eqref{eq:Lef} works for
any $T$ and $\tau$.

\subsubsection{When the DDP method succeeds}

For $T>\tau$ where the DDP method works, we shall quantitatively
justify our proposed method compared with the DDP method.  The
transition probability from the DDP method of Eq.~\eqref{eq:DDP} is
given by
\begin{equation}
  \label{eq:DDPmLZ}
  P \simeq \exp\left[-\frac{2T\Lambda}{\tau}
    E\left(\frac{\tau}{T},\frac{T}{\tau}\right)\right]\,,
\end{equation}
where $E(x,k)$ is the incomplete elliptic integral of the second kind.
We note that the above expression from the DDP method approaches the
exact analytical Landau-Zener formula, $P\to e^{-\pi\Lambda/2}$, as
$T\to\infty$.

\begin{figure}
  \centering
  \includegraphics[width=0.45\textwidth]{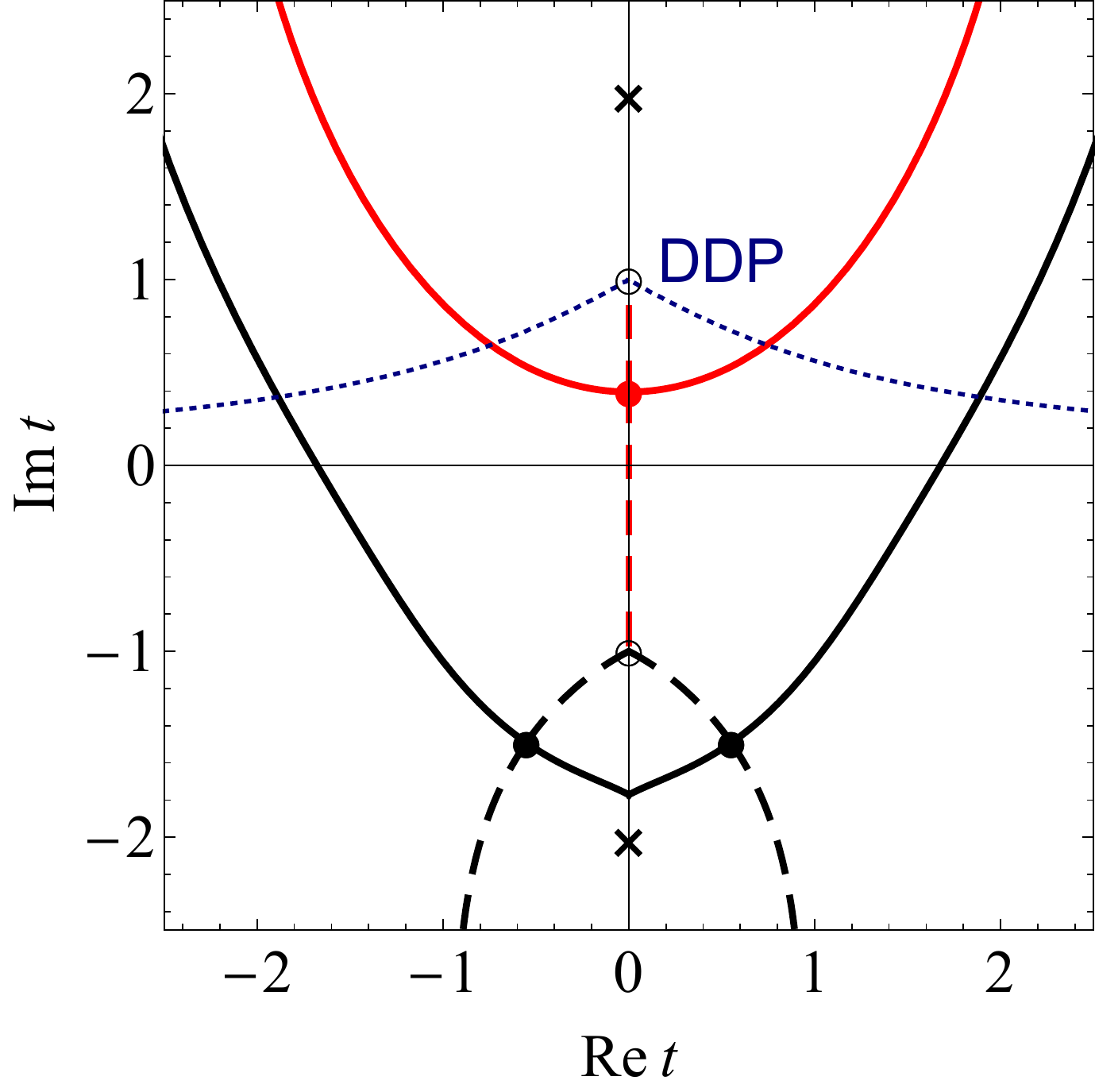}
  \caption{Analytical structure of $F(t)$ in complex-$t$ plane in the
    modified Landau-Zener model with $(T,\tau,\Lambda)=(2,1,1)$.  The
    upper (and lower) solid line represents the steepest descent with
    $n=1$ (and $n=0$, respectively).  The dashed lines are the
    steepest ascents and the dotted line is the DDP contour.  The
    cross dots represent the poles, the open-circle dots indicate the
    closing points, and the filled-circle dots are the saddle points.}
  \label{fig:SLSthimblesT=2}
\end{figure}

In the following, we address results for $(T,\tau,\Lambda)=(2,1,1)$
(for which $T > \tau$ is satisfied and the DDP method works) from a
new method we propose by Eq.~\eqref{eq:Lef}.  In
Fig.~\ref{fig:SLSthimblesT=2} we show the analytical structure of
$F(t)$ read from Eq.~\eqref{eq:mLZ} with the steepest descents and
ascents.  We implicitly assume some scale $t_0$ of $t$ and measure $T$
and $\tau$ in units of $t_0$ [where $t_0$ finally goes away, see
Eq.~\eqref{eq:DDPmLZ}].  In Fig.~\ref{fig:SLSthimblesT=2} the blue
dotted line represents the DDP contour attached to the closing point
$t_c$ shown by the open-circle dots.  It is clear that the poles shown
by the cross dots are harmless for the DDP contour deformation in this
case.  Three filled-circle dots denote the saddle points $t_s$ of
$F(t)$.  Solid and dashed lines represent the steepest descents and
ascents, respectively.  Only the upper (red) steepest ascent crosses
the original integration contour (namely, the real axis), so that the
upper (red) steepest descent has $n=1$ and contributes to recover the
original integration.  Two saddle points and one steepest descent
(black solid line) in ${\rm Im}\,t<0$ region are completely irrelevant
since the associated Morse index is vanishing.  We make a brief remark
on branch cuts on Fig.~\ref{fig:SLSthimblesT=2}.  Our choice of branch
cuts runs from $t=\pm i\tau, \pm iT$ along the imaginary axis in such
a way that they never cross the real axis.

\begin{figure}
  \centering
  \includegraphics[width=0.6\textwidth]{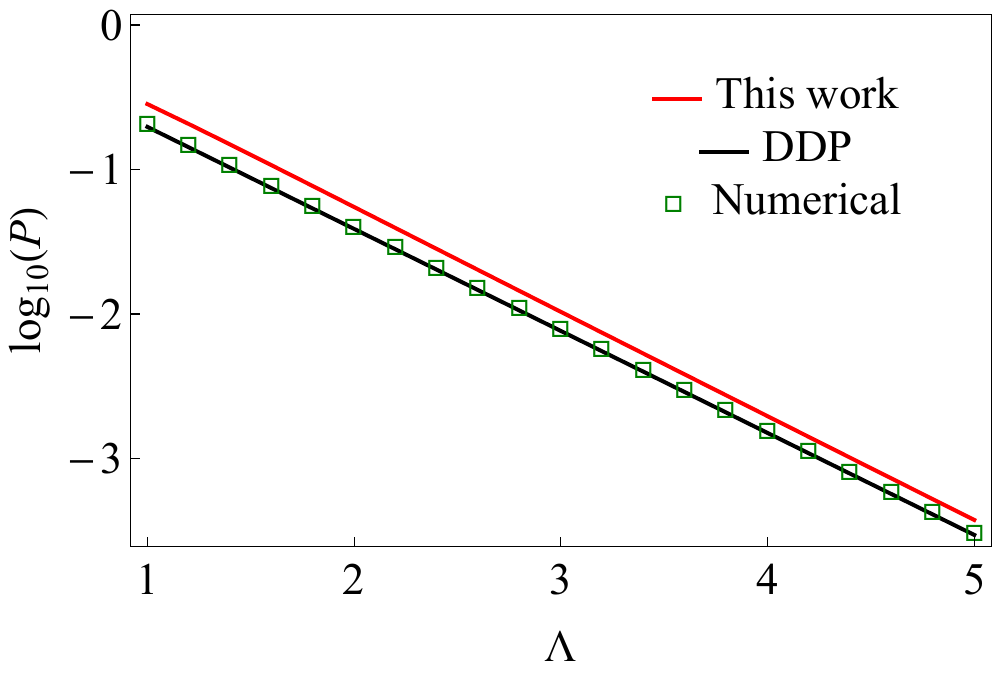}
  \caption{Comparison of $P$ as a function of $\Lambda$ in the
    modified Landau-Zener model with $(T,\tau)=(2,1)$.  The upper (and
  lower) solid line represents the results from this work (and the DDP
  method, respectively).  The open-square dots represent the full numerical
  results without truncation.}
\label{fig:SLSprobT=2}
\end{figure}

For this case of $(T,\tau)=(2,1)$ Fig.~\ref{fig:SLSprobT=2} presents
$P$ as a function of $\Lambda$;  the red solid line represents the
estimate from our new method, the black solid line shows the DDP
results, and the green open-square dots are from the direct numerical
calculation (without truncation).

One might think that the DDP method shows good agreement with
numerical calculations.  In fact such good agreement is quite
unnatural and accidental;  the DDP method relies on the first order
truncation in the iterative approximation, while the numerical
calculations shown here do \textit{not}.  Therefore, this level of
agreement exceeds theoretical limitation.  The point is that the DDP
method artificially reproduces the exact analytical formula for the
Landau-Zener model in the limit of $T\to\infty$, and so the deviation
is still negligibly small even at $T=2$.  Thus, we must conclude that
this too good agreement of the DDP method is due to a rather
accidental reason.

In contrast, the Lefschetz-thimble inspired method is mathematically
founded on the Picard-Lefschetz theory.  We should emphasize that the
results from our method in Fig.~\ref{fig:SLSprobT=2}, especially the
exponential slope, is reasonably consistent with the numerical
calculations within approximation uncertainties.  It is not very
surprising that the exponents from the DDP method and ours are such
close.  Actually, the exponent is dominated by the contribution from
$\Delta(t)$, and a shift from $\eta(t)$ affects only the prefactor.
On Fig.~\ref{fig:SLSthimblesT=2} we see that in ${\rm Im}\,t>0$ region
the open-circle dot ($t_c$) and the filled-circle dot ($t_s$) are
located differently, and their spacing is given by the effect of
$\eta(t)$.  More interestingly, in Fig.~\ref{fig:SLSthimblesT=2},
the blue dotted line denotes the DDP deformed contour [on which
${\rm Im}\,\Delta(t)={\rm Im}\,\Delta(t_c)$], and this is completely
different from the red solid line of our contour [on which
${\rm Im}\,F(t)=\text{(const.)}$].  According to our knowledge on the
Picard-Lefschetz theory there is no strict reason why the blue dotted
line of the DDP contour should be favored for the evaluation of
Eq.~\eqref{eq:aplus}.

\subsubsection{When the DDP method fails}

\begin{figure}
  \centering
  \includegraphics[width=0.45\textwidth]{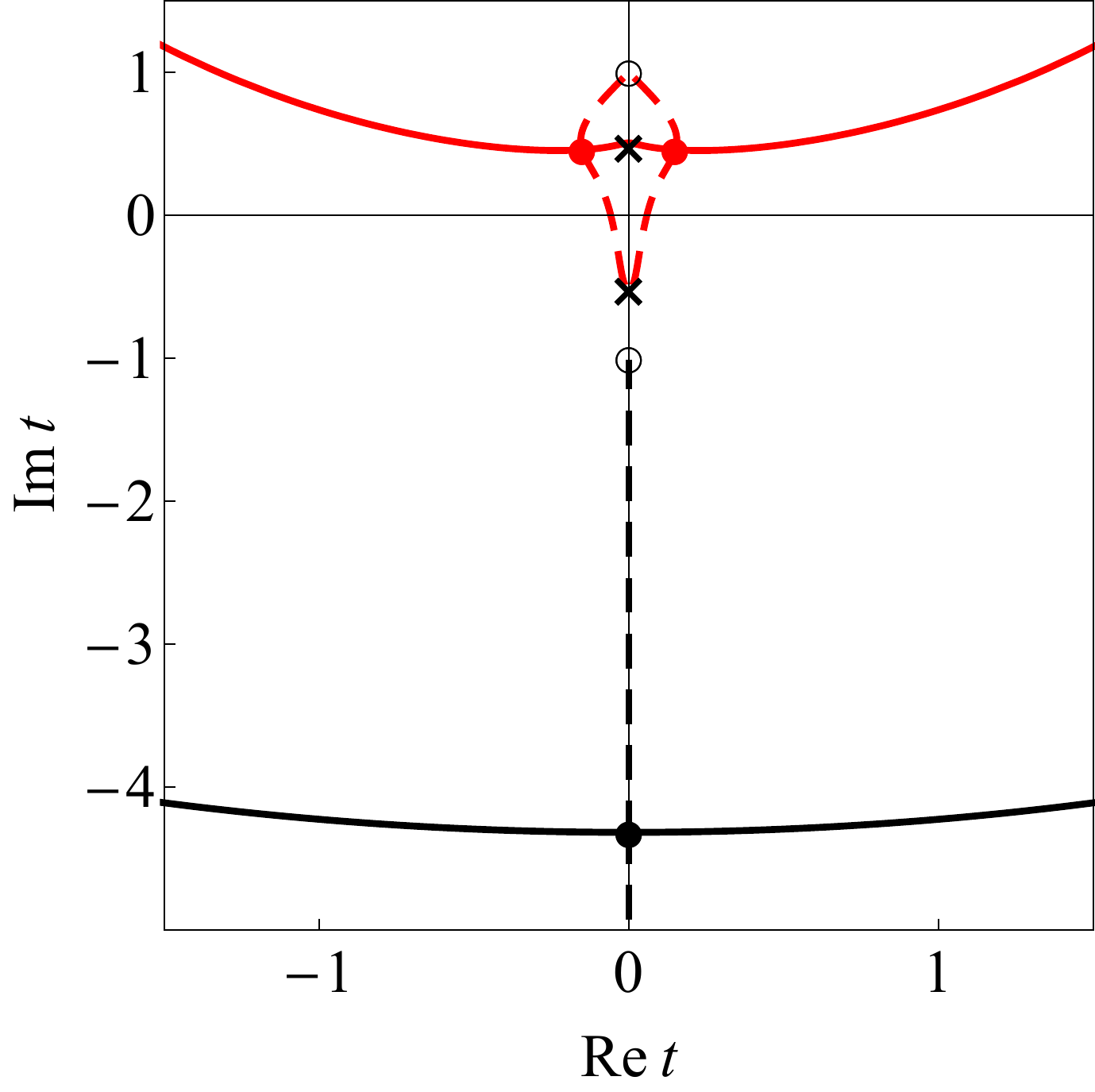}
  \caption{Analytical structure of $F(t)$ in complex-$t$ plane in the
    modified Landau-Zener model with $(T,\tau,\Lambda)=(\frac{1}{2},1,1)$.
    The upper (and lower) solid line represents the steepest descent
    with $n=1$ (and $n=0$, respectively).  The dashed lines are the
    steepest ascents.  The cross dots represent the poles, the
    open-circle dots indicate the closing points, and the
    filled-circle dots are the saddle points.}
  \label{fig:SLSthimblesT=1/2}
\end{figure}

For $T \leq \tau$ the DDP method is inapplicable due to the pole at
$t=iT$, whereas our method is still operative.  It is very interesting
to see how our method circumvents the pole in complex-$t$ plane.  In
Fig.~\ref{fig:SLSthimblesT=1/2} we show the analytical structure of
$F(t)$ with the steepest descents and ascents for
$(T,\tau,\Lambda)=(\frac{1}{2},1,1)$, for which $T<\tau$ and the DDP
method does not work.  Two filled-circle dots of the saddle points in
${\rm Im}\,t>0$ region have the steepest ascents crossing the real
axis, and so the associated steepest descents contribute to the
integral, i.e., $n=1$.  As mentioned before, we choose branch cuts
along the imaginary axis to avoid crossing the real axis.  In
Fig.~\ref{fig:SLSthimblesT=1/2}, strictly speaking, the black saddle
point on the branch cut is degenerated, but such an analytical
structure is irrelevant, for these saddle points make no contribution
in our approach (i.e., the associated $n=0$).  Here, the contour
modifications induced by $\eta(t)$ play an important role to avoid
the pole.  As mentioned before, the open-circle dots are changed to
the filled-circle does due to $\eta(t)$.  In
Fig.~\ref{fig:SLSthimblesT=1/2} we notice that two thimbles appear
associated with two saddle points, and one edge of each is attached to
the pole.  Thus, the pole is not crossed over, so that the original
integral is recovered exactly.  Such analytical structures hold for
general $T<\tau$.

\begin{figure}
  \centering
  \includegraphics[width=0.65\textwidth]{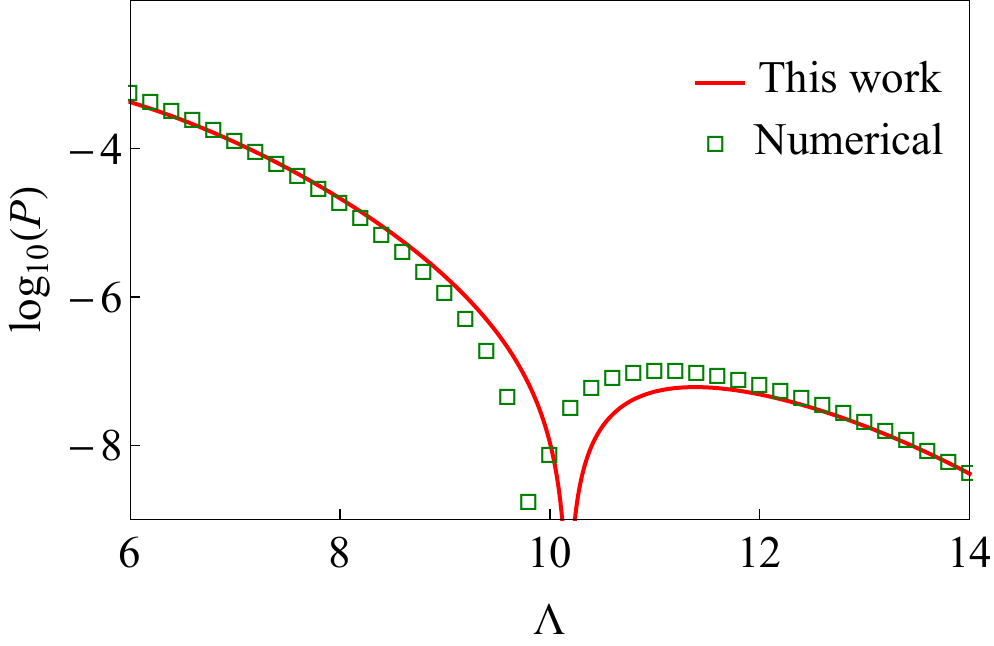}
  \caption{Comparison of $P$ as a function of $\Lambda$ in the
    modified Landau-Zener model with $(T,\tau)=(\frac{1}{2},1)$.  The
    solid line represents the results from this work and the
    open-square dots represent the numerical results.}
  \label{fig:SLSprobT=1/2}
\end{figure}

Figure~\ref{fig:SLSprobT=1/2} presents $P$ for
$(T,\tau)=(\frac{1}{2},1)$ as a function of $\Lambda$; the red solid
line represents the estimate from our new method and the green
open-square dots are from the full numerical calculation.  There is no
data from the DDP method that simply does not work.  We see a sharp
suppression of the probability around $\Lambda\simeq10$.  This
singular behavior is understandable from the similarity between the
modified Landau-Zener and the Rosen-Zener
models~\cite{rosen1932double}.  Our present work agrees fairly well
with the numerical calculation except for the region in the vicinity
of $\Lambda\simeq10$.  One possible explanation for this discrepancy
around $\Lambda\simeq10$ lies in the first order truncation
approximation to set $a_{-}(t)=1$ when we derived
Eq.~\eqref{eq:aplus}.   In the region where $P$ is vanishingly small,
higher order terms from multiple transitions between
$\ket{\chi_{\pm}(t)}$ are no longer negligible.

\begin{figure}
  \centering
  \includegraphics[width=0.65\textwidth]{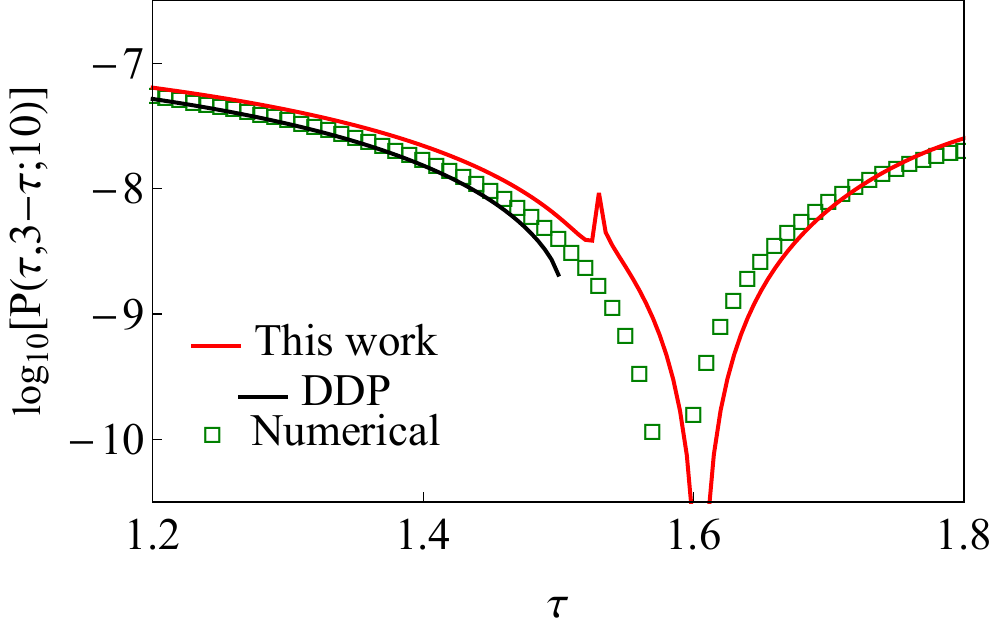}
  \caption{2D cross section of $P(\tau,T;\Lambda=10)$ cut at
    $T=3-\tau$.  The upper (and lower) solid line represents the
    results from this work (and the DDP method, respectively).  The
    open-square dots represent the numerical results.}
  \label{fig:resultd2}
\end{figure}

Now, we put our emphasis on the robustness of our new method compared
with the DDP method.  The probability, $P(\tau,T;\Lambda)$, is a function of
$T$, $\tau$, and $\Lambda$.  We could have shown a 3D plot of
$P(\tau,T;\Lambda)$ for a fixed $\Lambda$ but, here, we plot
$P(\tau,3-\tau;10)$ as a function of $\tau$ as shown in
Fig.~\ref{fig:resultd2}.  This is a 2D cross section of $P(\tau,T;10)$
cut by $T=3-\tau$ along which we can see both regions where the DDP
method does and does not work.  For $\tau \geq 1.5$ the DDP method
is inoperative, on the one hand, so the black line ends at $\tau=1.5$ in
Fig.~\ref{fig:resultd2}.  Our method, on the other hand, gives
consistent estimates for any $\tau$, including the parameter region of
$[(T,\tau,\Lambda)|T\leq\tau]$ which the DDP method fails to access.

Let us make a couple of comments on Fig.~\ref{fig:resultd2}.  Firstly,
we notice a peak around $\tau\simeq1.55$ in this work.  This peak is
caused by the fact that the structure of the steepest descents
drastically changes at $\tau\simeq1.55$.  In fact, the analytical
structure changes from one like Fig.~\ref{fig:SLSthimblesT=2} to the
other like Fig.~\ref{fig:SLSthimblesT=1/2}.  Accordingly one critical
point that contributes to the integral splits into two.  These two
critical points are close to each other shortly after the splitting,
so the Gaussian approximation, as done in Eq.~\eqref{eq:Lef}, becomes
unreliable around $\tau\simeq1.55$.  Secondly, we again see a sharp
suppression of the probability around $\tau\simeq1.6$, as in
Fig.~\ref{fig:SLSprobT=1/2}.  The underlying mechanism for this
suppression is identical with the previous discussions.
Overall, we can conclude that our new method supersedes the DDP
method.

\section{Schwinger Mechanism}

As another interesting application let us discuss the Schwinger
Mechanism.  One might have an impression that the pair production
problem in Quantum Field Theories appears to be distinct from
two-level quantum mechanical problems.  As correctly pointed out by
Cohen and McGady~\cite{Cohen:2008wz}, however, the Dirac equation for
the Schwinger Mechanism reduces to the same form as the Landau-Zener
model.  We can immediately make sure that the exponential factor well
known for the Schwinger Mechanism is readily derived from the DDP
method, as sketched below.  Then, a further challenging question is
when the DDP method breaks down and how our method finds the correct
answer.

\begin{figure}
  \centering
  \includegraphics[width=0.48\textwidth]{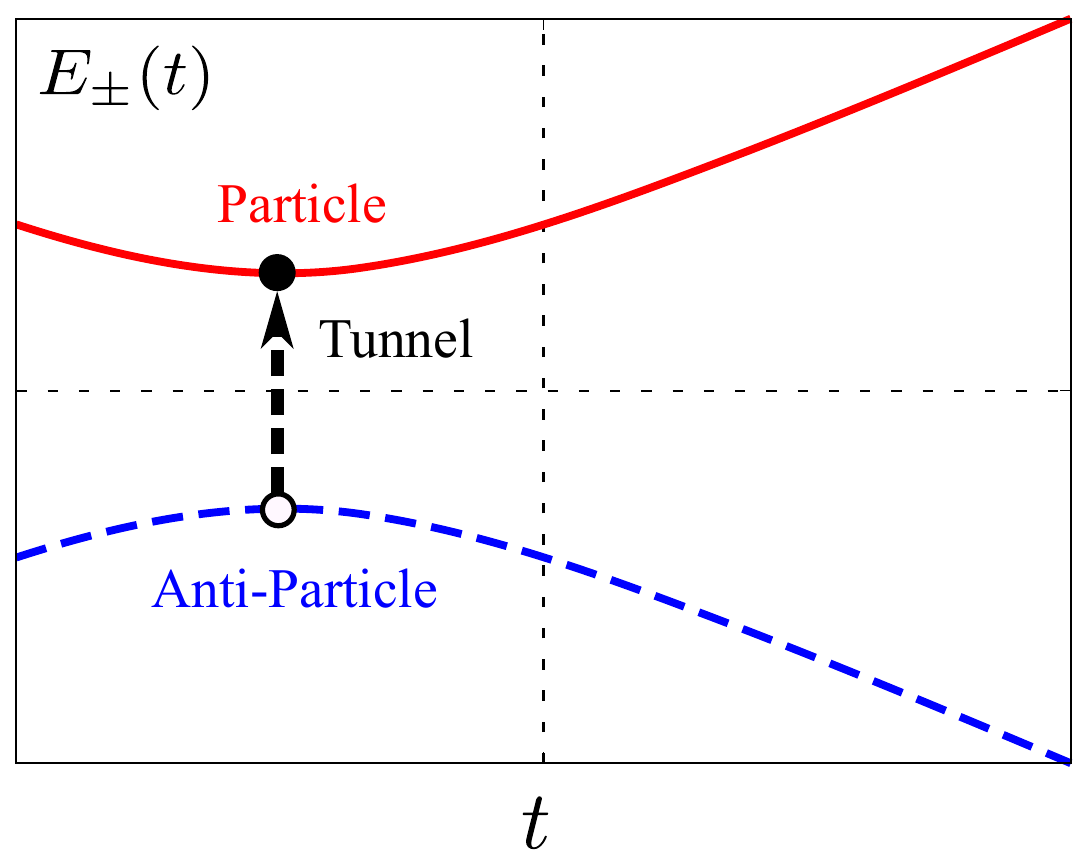}
  \caption{Particle and anti-particle energy levels for a constant
    electric field.  The pair production process corresponding to
    quantum tunneling from the anti-particle to the particle state.}
  \label{fig:schwinger}
\end{figure}

\subsection{Constant electric field}

The pair production process is nothing but a transition from the
anti-particle state (with negative energy) to the particle state (with
positive energy) as sketched in Fig.~\ref{fig:schwinger}.  In this
paper we limit ourselves to systems with translational invariance.
Then, the spatial momenta are simply labels for eigenenergies and
there will be no mixing. More specifically, if the external electric
field is spatially homogeneous and along the $z$ axis, the system is
effectively described by the following Hamiltonian;
\begin{equation}
  H(t) = \left[
  \begin{array}{cc}
    p_z - eA_z(t) & m_\perp \\
    m_\perp & -p_z + eA_z(t)
  \end{array}
  \right]\,,
  \label{eq:hamiltonian_sch}
\end{equation}
where we introduced $m_\perp=\sqrt{p_x^2+p_y^2+m^2}$.  This form of
the Hamiltonian corresponds to the Landau-Zener model with $\alpha(t)$
and $V(t)$ in Eq.~\eqref{eq:hamiltonian} chosen as
\begin{equation}
  \alpha(t) = 2[p_z - eA_z(t)]\,, \qquad
  V(t) = 2m_\perp\,.
\end{equation}
In the case of the constant electric field, $A_z(t) = -Et$, two
adiabatic eigenenergies, namely, two eigenvalues of the
matrix~\eqref{eq:hamiltonian_sch} are
\begin{equation}
    E_{\pm}(t) = \pm \sqrt{ (p_z + eE t)^2 + m_\perp^2 }\,.
\end{equation}
We can also calculate the adiabatic factor, $\eta(t)$,
which plays a pivotal role in our method.
For this problem, obviously, there is no pole in $\delta E(t)$ and
the DDP method works. We assume below that the mass gap is larger
than the magnitude of the electric field, $m^2 \gg eE$.

The particle and the anti-particle levels do not touch as long as $t$
is real, as is obvious in Fig.~\ref{fig:schwinger}.  In the
complex-$t$ plane we can locate the closing point simply from the
condition, $E_{\pm}(t=t_c)=0$, from which we find,
\begin{equation}
  t_c = -\frac{p_z}{eE} + \rmi \frac{m_\perp}{eE}\,.
  \label{eq:tc}
\end{equation}
We can say that the classically prohibited Schwinger process becomes
possible for $t$ to take such a detour to $t_c$ where two states
touch.  According to Eq.~\eqref{eq:DDP} we can immediately find the
exponent as
\begin{align}
    &-2{\rm Im}\,\Delta (t_c)
    = -4{\rm Im}\, \int_0^{t_c} \rmd t\, \sqrt{(p_z + eE t)^2 + m_\perp^2} \notag\\
    &\qquad = -\frac{2}{eE}{\rm Im}\, \Biggl\{
     (p_z + eE t) \sqrt{(p_z + eE t)^2 + m_\perp^2} + m_\perp^2
     \ln\Bigl[ (p_z + eE t) + \sqrt{(p_z + eE t)^2+m_\perp^2} \Bigr]
     \Biggr\}_{t=0}^{t=t_c} \notag\\
    &\qquad = -\frac{2m_\perp^2}{eE} {\rm Im}[ \ln(\rmi ) ]
    = -\frac{\pi m_\perp^2}{eE} \,.
\end{align}
Thus, we can approximate the transition probability,
$P\simeq \rme^{-\pi m_\perp^2 / (eE)}$, as is perfectly consistent
with the well-known Schwinger formula.  We note that the prefactor for
the $(3+1)$-dimensional theory should be retrieved from the transverse
phase-space integration.  This is a successful example to show how the
DDP method is useful for a wide class of physics problems even in
Quantum Field Theories.

\subsection{Pulse electric field}

A less successful example of the DDP formula application is soon
encountered once the electric field profile gets time dependent.  The
analytical properties of the Sauter-type potential are well understood
(see Ref.~\cite{Dunne:2004nc} and references therein), which is given
by
\begin{equation}
  e A_z (t) = -\frac{eE}{\omega} \tanh(\omega t) \,,
  \label{eq:sauter}
\end{equation}
leading to a pulse-shaped profile of electric field, i.e.,
\begin{equation}
    eE(t) = \frac{eE}{\cosh^2(\omega t)} \,.
\end{equation}
Now, for later convenience, let us introduce the Keldysh adiabaticity
parameter as done in Ref.~\cite{Schutzhold:2008pz} as
\begin{equation}
  \gamma = \frac{m\omega}{eE}\,.
  \label{eq:gamma}
\end{equation}
The electric profile smoothly reduces to a constant electric field in
the limit of $\omega\to 0$ or $\gamma\to 0$.
The Hamiltonian describing this system is Eq.~\eqref{eq:hamiltonian_sch} with
Eq.~\eqref{eq:sauter}. The energy difference in this case is,
\begin{equation}
  \delta E(t) = 2\sqrt{\Bigl[ p_z
    + \frac{m}{\gamma}\tanh(\omega t) \Bigr]^2 + m_\perp^2} \,.
\end{equation}
The adiabatic factor, $\eta(t)$, is calculable straightforwardly as well.
We again emphasize the importance of $\eta(t)$ in comparing between the DDP method and our method. The closing points are located at
\begin{equation}
    t_c = \frac{1}{\omega}\tanh^{-1} \biggl( -\frac{\gamma p_z}{m}
    + \rmi \frac{\gamma m_\perp}{m} \biggr)
    + \rmi \frac{2\pi k_1}{\omega}
\end{equation}
for $k_1\in \mathbb{Z}$.  Because contributions from $k_1\neq 0$ are
suppressed by large exponents, it would be sufficient for us to
consider only the $k_1=0$ contribution.  For this specific choice of
the vector potential, there are poles at
\begin{equation}
  t_{\text{pole}} = \rmi \frac{\pi (1+2k_2)}{2\omega}
  \label{eq:sauter_pole}
\end{equation}
with $k_2\in \mathbb{Z}$.  Now, let us assume a sufficiently small
value of $\omega$ so that ${\rm Im}\,t_{\text{pole}} \ll {\rm Im}\,t_c$
and thus the DDP method definitely works.  Under this condition the
DDP method gives us an approximated expression, $P\simeq \rme^{-A m^2/(eE)}$,
with
\begin{equation}
    A = \frac{\pi}{\gamma m} \Biggl[
      \sqrt{\Bigl(\frac{m}{\gamma} + p_z\Bigr)^2 + m_\perp^2}
    + \sqrt{\Bigl(\frac{m}{\gamma} - p_z\Bigr)^2 + m_\perp^2}
    - \frac{2m}{\gamma}  \Biggr] \,,
\label{eq:DDP_sauter}
\end{equation}
which can be further simplified into
\begin{equation}
   A \simeq \frac{\pi m_\perp^2}{2m} \Biggl(
    \frac{1}{m+\gamma p_z} + \frac{1}{m-\gamma p_z} \Biggr)
\end{equation}
for $\gamma (p_z/m) \ll 1$.  This exactly agrees with the expanded
results from the analytical answer~\cite{Fukushima:2009er}.  Clearly,
we recover the previous result for constant electric field in the adiabatic
$\gamma\to 0$ limit.  It is also natural that $A\to \infty$ and thus
$P\to 0$ as $p_z$ approaches $\pm m/\gamma = \pm eE/\omega$ since
the maximum $p_z$ should be an impulse by a force $eE$ and a duration
$1/\omega$.  If $p_z$ exceeds this maximum, the pair production is
disfavored and $P\to 0$ follows.

In the opposite limit of $\gamma\gg 1$ the impulse approximation would
work better to justify the perturbative expansion.  However, from
Eq.~\eqref{eq:DDP_sauter}, $A\to 0$ and $P\to 1$ is concluded in the
$\gamma\to\infty$ limit.  In this case the DDP method does not give the
correct asymptotic behavior.  The reason why the DDP method does not
work is more nontrivial than the modified Landau-Zener model.  In
fact, with the Sauter-type electric field, the DDP contour is
deformable without hitting any pole in the complex-$t$ plane.  The
validity lost of the DDP method is attributed to the adiabatic factor
$\eta(t)$ which is not taken into account in the determination of
$t_c$.  Our results from the Lefschetz-thimble inspired method take
care of $\eta(t)$, so that they can be consistent with the full
answer.

\begin{figure}
  \centering
  \includegraphics[width=0.65\textwidth]{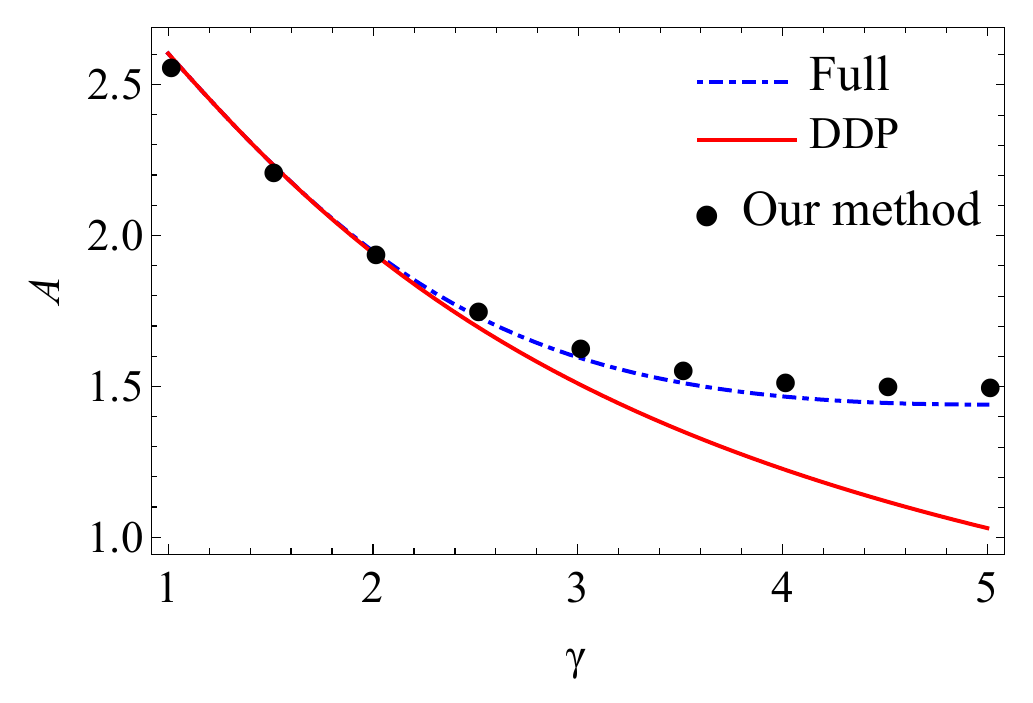}
  \caption{Comparison of $A$ of $P\simeq \rme^{-A m^2/(eE)}$ as a
    function of the Keldysh parameter $\gamma$ between the full answer
    (dotted line), the DDP method (solid line), and our estimates
    (filled circles) for $m=3$ and $eE=3$ at $\boldsymbol{p}=0$ in
    arbitrary unit.}
  \label{fig:sauter}
\end{figure}

Figure~\ref{fig:sauter} presents a quantitative comparison between the
full answer (dotted line), the DDP method (solid line), and our
estimates (filled circles).  The DDP method is a pretty good
approximation for $\gamma \lesssim 2$, but it loses the agreement with
the full answer for larger $\gamma$.  In contrast, owing to the
inclusion of $\eta(t)$, our method gives reasonablly approximating
results for overall $\gamma$.

\subsection{Dynamically Assisted Schwinger Mechanism}

Finally, let us consider an example which has attracted a lot of
theoretical attentions.  That is, we will consider a combination of a
constant electric field and a pulsed one, which is represented by the
following vector potential,
\begin{equation}
  eA_z(t) = -eEt - \frac{e\varepsilon}{\omega}\tanh(\omega t)\,,
\end{equation}
where $\varepsilon/E \ll 1$ is presumed.  In the same way as the
previous subsection, we can immediately obtain the adiabatic energy
difference,
\begin{equation}
  \delta E(t) = 2\sqrt{\Bigl[ p_z
    + eEt + \frac{e\varepsilon}{\omega}\tanh(\omega t) \Bigr]^2 + m_\perp^2} \,.
\end{equation}
The cloing points are the solutions of
\begin{equation}
  \omega t_c + \frac{\varepsilon}{E}\tanh(\omega t_c)
  = -\frac{\gamma p_z}{m} + \rmi \frac{\gamma m_\perp}{m} \,.
\end{equation}
Here, we should note that we share the same notation, $\gamma$, as
defined in Eq.~\eqref{eq:gamma}, but the meaning of $E$ and $\omega$
in this subsection are different from the previous subsection;
$\omega$ is a frequency of the field whose strength is not $E$ but
$\varepsilon$ now.  The pair production at $\boldsymbol{p} = 0$ has the least
exponential suppression, so we set $\boldsymbol{p} = 0$ for simplicity below.
This choice simplifies the location of the closing points and enables
us to perform analytic calculation to some extent.  The closing
points, $t_c = i \tau_c$, are the solutions of
\begin{equation}
  \omega \tau_c + \frac{\varepsilon}{E}\tan(\omega \tau_c) = \gamma\,.
  \label{eq:dasmclosing}
\end{equation}
Although $\varepsilon/E \ll 1$, the second term in the left-hand side
of Eq.~\eqref{eq:dasmclosing} is not negligible because of the
singularities from $\tan(\omega \tau_c)$.  The above singularities
also give the poles of $\delta E(t)$, i.e.,
\begin{equation}
  t_{\text{pole}} = \rmi \frac{\pi(1 + 2k)}{2\omega}
\end{equation}
with $k \in \mathbb{Z}$ just in the same manner as in
Eq.~\eqref{eq:sauter_pole}.

The solutions of Eq.~\eqref{eq:dasmclosing} can be easily drawn by
the crossing points between two functions, $(\varepsilon/E)\tan(x)$
and $-x+\gamma$, where $x=\omega\tau_c$.  Because of the divergent
behavior of $\tan(x)$, the closing point nearest to the origin in the
upper complex-$t$ plane is $\omega \tau_c=\pi/2 - \delta \,(\delta\ll 1)$
for $\gamma \gtrsim \pi/2$.  Therefore, for sufficiently small
$\gamma \ll 1$, we see that $t_{\text{pole}}$ does not intrude the
region of the contour deformation toward $t_c=i\tau_c$, and the DDP method
should work.  Indeed, the DDP method gives us the asymptotic form of
the probability, $P \simeq \rme^{-Am^2/(eE)}$, with
\begin{equation}
  A = \frac{2\pi}{\gamma}
\end{equation}
for $\gamma \gg 1$.  This asymptotic behavior inferred from the DDP
formula agrees with the result from the worldline
instanton~\cite{Schutzhold:2008pz}.  It should be noted that the
direct comparison is little subtle;  we computed the rate for $\boldsymbol{p}=0$
for simplicity, and the exponent (or the vacuum decay probability)
obtained in Ref.~\cite{Schutzhold:2008pz} is a quantity after the
whole phase-space integration.  As long as we focus on the exponent
only, the quantitative comparison still makes sense since the $\boldsymbol{p}=0$
contribution is dominant.

\begin{figure}
  \centering
  \includegraphics[width=0.65\textwidth]{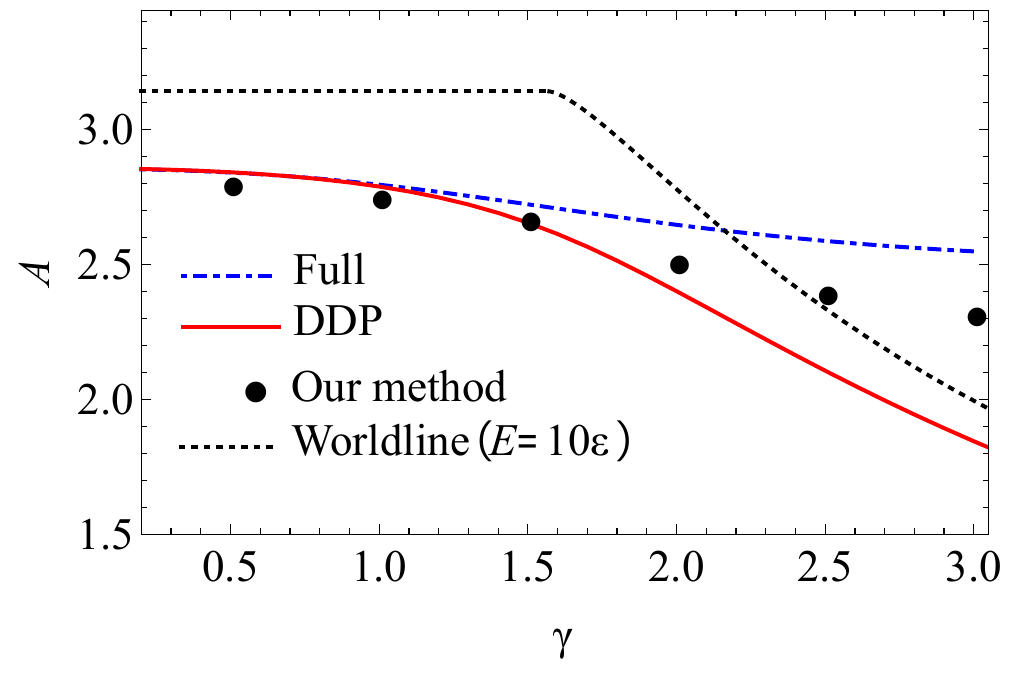}
  \caption{Comparison of $A$ of $P\simeq \rme^{-A m^2/(eE)}$ as a
    function of the Keldysh parameter $\gamma$ between the full
    solution (dotted line), the DDP method (solid line), and our
    estimates (filled circles) for $m=3$, $eE=3$, and
    $\varepsilon=0.3$ at $\boldsymbol{p}=0$ in arbitrary unit.}
  \label{fig:dasm}
\end{figure}

When we derived the asymptotic form of the DDP formula, we only
imposed $\gamma \gg 1$ and $E \gg \varepsilon$.  However, another condition,
$m \gg \omega$, is required to justify the worldline instanton
calculus in order to prohibit the dynamical pair production.
Figure.~\ref{fig:dasm} presents a detailed comparison between the full
answer (dot-dashed line), the DDP method (solid line), the worldline
method (dotted line), and our estimate (filled circles) when we chose
$m=3,\,eE=3$ and $\varepsilon = 0.3$.  We can confirm that the DDP and
the worldline results approach each other for large $\gamma$, as is
consistent with the analytical consideration.  A surprise comes from
the fact that the full answer from the brute-force numerical
calculation deviates from the DDP and the worldline results, which is
to be explained by the breakdown of $m \gg \omega$ for large
$\gamma$ with the current parameter set.  Our results from the
Lefschetz-thimble inspired method tend to stay closer to the full
answer than the other methods.  The improvement is again ascribed to
the adiabatic factor $\eta(t)$ which is taken into account in our
method.

\section{Conclusions}

We studied the Dykhne-Davis-Pechukas (DDP) method to describe a
transition amplitude in two-level quantum systems and proposed an
alternative method inspired by the Lefschetz-thimble method.  Since
the DDP method is based on complex deformation of the time
integration, one may think that the DDP method is justified directly
from the complex analysis.  Contrary to that, howevever, the one
dimensional application of the Picard-Lefschetz theory implies that
the optimal contour is different from the one adopted in the DDP
method.

With some parameters of the Hamiltonian for which the DDP complex-$t$
contour goes across poles and/or the closing points are too close to
poles, the DDP method breaks down.  Our proposed method can evade this
problem thanks to the pole-free structure of the steepest descents
whose shape is affected by poles.  We explicitly tested the perfomance
of our method in the modified Landau-Zener model.  The comparison
between the direct numerical calculation, the DDP method, and our
method confirmed that our method can give a reasonable approximation
for all parameter regions even when the DDP method cannot.

Two-level quantum systems include interesting modeling of quantum
field-theoretical problems.  Among them, we treated the Schwinger
Mechanism, i.e., the pair production of a particle and an
anti-particle under electric fields, as a two-level quantum problem in
the Landau-Zener model.  For the constant electric field the DDP
method gave an answer fully consistent with the known formula.  We
found, however, that the approximation of the DDP method gets worse
for pulsed shapes of electric fields if the pulse is short-lived and
the adiabatic factor is not negligible for large Keldysh factor
$\gamma>2$.  In contrast our method gives results in agreement with
the full numerical answer for parameters when the DDP method loses
validity.  We further applied these methods for the Dynamically
Assisted Schwinger Mechanism in which the exponential threshold is
suppressed.  Our calculations show that both the DDP method and our
method match well with the full answer for small $\gamma$ regions.
The interesting is for larger $\gamma>2$ where the DDP method ceases
working well.  The DDP method leads to an exponent that asymptotically
approaches the value obtained in the worldline instanton method as
$\gamma\to \infty$ (in which the worldline instanton calculus is
anyway out of validity region).  In our method the adiabatic factor is
taken into account and the approximated estimate turns out to be
closer to the full answer than other methods.

There are various future directions to extend the current work.  One
of them lies in a closer look at the foundation of the DDP method.
The Picard-Lefschetz theory cannot justify the DDP method, but it is
certainly true that the DDP method somehow reproduces the exact
analytical results for the Landau-Zener model and there might be some
underlying mechanisms for such a coincidence.  Besides, as we found,
the DDP method seems to be compatible with the worldline instanton
method, and its reasoning would be, if any, quite interesting to seek
for.

Finally, we shall mention an intriguing possibility for future works.
In addition to advantages in a sense of adaptivity of our method, it
could be the case that our method can be a new tool to reveal some
drastic changes of model features like a phase transition.  It is
actually known that the structure of the steepest descents or the
Lefschetz thimbles in more general may suddenly change, which is
referred to as the Stokes phenomenon.  In the examples taken in this
work, there was not such drastic and sudden alteration in the
structure of the steepest descents, but if we consider a wider class
of physics problems such as the Floquet-type problems with time
periodic electric disturbances, some transitional changes would be
caused.  Once the Stokes phenomenon is involved, our method would be
far superior to other methods.  Our new method awaits phenomenological
applications in the band theory of solids as well as in quantum
field-theoretical problems such as the Schwinger Mechanism with more
general time-dependent electric/magnetic profiles, the Hawking
radiation on top of various metrics for which the analytical Bogoliubov
coefficients are not calculable, etc, which will be reported elsewhere.

\section*{acknowledgments}
The authors thank
Takashi~Oka
and
Kazuaki~Takasan
for stimulating comments and discussions.
This work was supported by Japan Society for the Promotion of Science
(JSPS) KAKENHI Grant No.\ 18H01211 and 19K21874.

\bibliography{DDP}

\begin{thebibliography}{50}%
\makeatletter
\providecommand \@ifxundefined [1]{%
 \@ifx{#1\undefined}
}%
\providecommand \@ifnum [1]{%
 \ifnum #1\expandafter \@firstoftwo
 \else \expandafter \@secondoftwo
 \fi
}%
\providecommand \@ifx [1]{%
 \ifx #1\expandafter \@firstoftwo
 \else \expandafter \@secondoftwo
 \fi
}%
\providecommand \natexlab [1]{#1}%
\providecommand \enquote  [1]{``#1''}%
\providecommand \bibnamefont  [1]{#1}%
\providecommand \bibfnamefont [1]{#1}%
\providecommand \citenamefont [1]{#1}%
\providecommand \href@noop [0]{\@secondoftwo}%
\providecommand \href [0]{\begingroup \@sanitize@url \@href}%
\providecommand \@href[1]{\@@startlink{#1}\@@href}%
\providecommand \@@href[1]{\endgroup#1\@@endlink}%
\providecommand \@sanitize@url [0]{\catcode `\\12\catcode `\$12\catcode
  `\&12\catcode `\#12\catcode `\^12\catcode `\_12\catcode `\%12\relax}%
\providecommand \@@startlink[1]{}%
\providecommand \@@endlink[0]{}%
\providecommand \url  [0]{\begingroup\@sanitize@url \@url }%
\providecommand \@url [1]{\endgroup\@href {#1}{\urlprefix }}%
\providecommand \urlprefix  [0]{URL }%
\providecommand \Eprint [0]{\href }%
\providecommand \doibase [0]{http://dx.doi.org/}%
\providecommand \selectlanguage [0]{\@gobble}%
\providecommand \bibinfo  [0]{\@secondoftwo}%
\providecommand \bibfield  [0]{\@secondoftwo}%
\providecommand \translation [1]{[#1]}%
\providecommand \BibitemOpen [0]{}%
\providecommand \bibitemStop [0]{}%
\providecommand \bibitemNoStop [0]{.\EOS\space}%
\providecommand \EOS [0]{\spacefactor3000\relax}%
\providecommand \BibitemShut  [1]{\csname bibitem#1\endcsname}%
\let\auto@bib@innerbib\@empty
\bibitem [{\citenamefont {Oka}\ and\ \citenamefont
  {Aoki}(2010)}]{oka2010dielectric}%
  \BibitemOpen
  \bibfield  {author} {\bibinfo {author} {\bibfnamefont {T.}~\bibnamefont
  {Oka}}\ and\ \bibinfo {author} {\bibfnamefont {H.}~\bibnamefont {Aoki}},\
  }\href {\doibase 10.1103/PhysRevB.81.033103} {\bibfield  {journal} {\bibinfo
  {journal} {Phys. Rev. B}\ }\textbf {\bibinfo {volume} {81}},\ \bibinfo
  {pages} {033103} (\bibinfo {year} {2010})}\BibitemShut {NoStop}%
\bibitem [{\citenamefont {Oka}(2012)}]{oka2012nonlinear}%
  \BibitemOpen
  \bibfield  {author} {\bibinfo {author} {\bibfnamefont {T.}~\bibnamefont
  {Oka}},\ }\href {\doibase 10.1103/PhysRevB.86.075148} {\bibfield  {journal}
  {\bibinfo  {journal} {Phys. Rev. B}\ }\textbf {\bibinfo {volume} {86}},\
  \bibinfo {pages} {075148} (\bibinfo {year} {2012})}\BibitemShut {NoStop}%
\bibitem [{\citenamefont {Berry}(2009)}]{berry2009transitionless}%
  \BibitemOpen
  \bibfield  {author} {\bibinfo {author} {\bibfnamefont {M.~V.}\ \bibnamefont
  {Berry}},\ }\href {\doibase 10.1088/1751-8113/42/36/365303} {\bibfield
  {journal} {\bibinfo  {journal} {Journal of Physics A: Mathematical and
  Theoretical}\ }\textbf {\bibinfo {volume} {42}},\ \bibinfo {pages} {365303}
  (\bibinfo {year} {2009})}\BibitemShut {NoStop}%
\bibitem [{\citenamefont {Sinitsyn}\ \emph {et~al.}(2018)\citenamefont
  {Sinitsyn}, \citenamefont {Yuzbashyan}, \citenamefont {Chernyak},
  \citenamefont {Patra},\ and\ \citenamefont {Sun}}]{sinitsyn2018integrable}%
  \BibitemOpen
  \bibfield  {author} {\bibinfo {author} {\bibfnamefont {N.~A.}\ \bibnamefont
  {Sinitsyn}}, \bibinfo {author} {\bibfnamefont {E.~A.}\ \bibnamefont
  {Yuzbashyan}}, \bibinfo {author} {\bibfnamefont {V.~Y.}\ \bibnamefont
  {Chernyak}}, \bibinfo {author} {\bibfnamefont {A.}~\bibnamefont {Patra}}, \
  and\ \bibinfo {author} {\bibfnamefont {C.}~\bibnamefont {Sun}},\ }\href
  {\doibase 10.1103/PhysRevLett.120.190402} {\bibfield  {journal} {\bibinfo
  {journal} {Phys. Rev. Lett.}\ }\textbf {\bibinfo {volume} {120}},\ \bibinfo
  {pages} {190402} (\bibinfo {year} {2018})}\BibitemShut {NoStop}%
\bibitem [{\citenamefont {Vasilev}\ \emph {et~al.}(2009)\citenamefont
  {Vasilev}, \citenamefont {Kuhn},\ and\ \citenamefont
  {Vitanov}}]{vasilev2009optimum}%
  \BibitemOpen
  \bibfield  {author} {\bibinfo {author} {\bibfnamefont {G.~S.}\ \bibnamefont
  {Vasilev}}, \bibinfo {author} {\bibfnamefont {A.}~\bibnamefont {Kuhn}}, \
  and\ \bibinfo {author} {\bibfnamefont {N.~V.}\ \bibnamefont {Vitanov}},\
  }\href {\doibase 10.1103/PhysRevA.80.013417} {\bibfield  {journal} {\bibinfo
  {journal} {Phys. Rev. A}\ }\textbf {\bibinfo {volume} {80}},\ \bibinfo
  {pages} {013417} (\bibinfo {year} {2009})}\BibitemShut {NoStop}%
\bibitem [{\citenamefont {Gu\'erin}\ \emph {et~al.}(2011)\citenamefont
  {Gu\'erin}, \citenamefont {Hakobyan},\ and\ \citenamefont
  {Jauslin}}]{guerin2011optimal}%
  \BibitemOpen
  \bibfield  {author} {\bibinfo {author} {\bibfnamefont {S.}~\bibnamefont
  {Gu\'erin}}, \bibinfo {author} {\bibfnamefont {V.}~\bibnamefont {Hakobyan}},
  \ and\ \bibinfo {author} {\bibfnamefont {H.~R.}\ \bibnamefont {Jauslin}},\
  }\href {\doibase 10.1103/PhysRevA.84.013423} {\bibfield  {journal} {\bibinfo
  {journal} {Phys. Rev. A}\ }\textbf {\bibinfo {volume} {84}},\ \bibinfo
  {pages} {013423} (\bibinfo {year} {2011})}\BibitemShut {NoStop}%
\bibitem [{\citenamefont {Zhang}\ \emph {et~al.}(2019)\citenamefont {Zhang},
  \citenamefont {Song}, \citenamefont {Ai}, \citenamefont {Wang}, \citenamefont
  {Yang},\ and\ \citenamefont {Deng}}]{zhang2019fast}%
  \BibitemOpen
  \bibfield  {author} {\bibinfo {author} {\bibfnamefont {H.}~\bibnamefont
  {Zhang}}, \bibinfo {author} {\bibfnamefont {X.-K.}\ \bibnamefont {Song}},
  \bibinfo {author} {\bibfnamefont {Q.}~\bibnamefont {Ai}}, \bibinfo {author}
  {\bibfnamefont {H.}~\bibnamefont {Wang}}, \bibinfo {author} {\bibfnamefont
  {G.-J.}\ \bibnamefont {Yang}}, \ and\ \bibinfo {author} {\bibfnamefont
  {F.-G.}\ \bibnamefont {Deng}},\ }\href {\doibase 10.1364/OE.27.007384}
  {\bibfield  {journal} {\bibinfo  {journal} {Opt. Express}\ }\textbf {\bibinfo
  {volume} {27}},\ \bibinfo {pages} {7384} (\bibinfo {year}
  {2019})}\BibitemShut {NoStop}%
\bibitem [{\citenamefont {Ma}\ \emph {et~al.}(2019)\citenamefont {Ma},
  \citenamefont {Chai}, \citenamefont {Liang}, \citenamefont {Duan},
  \citenamefont {Zhang}, \citenamefont {Dong},\ and\ \citenamefont
  {Shi}}]{ma2019high}%
  \BibitemOpen
  \bibfield  {author} {\bibinfo {author} {\bibfnamefont {R.-Q.}\ \bibnamefont
  {Ma}}, \bibinfo {author} {\bibfnamefont {B.-Y.}\ \bibnamefont {Chai}},
  \bibinfo {author} {\bibfnamefont {M.}~\bibnamefont {Liang}}, \bibinfo
  {author} {\bibfnamefont {Z.-L.}\ \bibnamefont {Duan}}, \bibinfo {author}
  {\bibfnamefont {W.-W.}\ \bibnamefont {Zhang}}, \bibinfo {author}
  {\bibfnamefont {J.}~\bibnamefont {Dong}}, \ and\ \bibinfo {author}
  {\bibfnamefont {J.}~\bibnamefont {Shi}},\ }\href {\doibase
  https://doi.org/10.1016/j.optcom.2018.08.027} {\bibfield  {journal} {\bibinfo
   {journal} {Optics Communications}\ }\textbf {\bibinfo {volume} {430}},\
  \bibinfo {pages} {1 } (\bibinfo {year} {2019})}\BibitemShut {NoStop}%
\bibitem [{\citenamefont {Sauter}(1931)}]{Sauter:1931zz}%
  \BibitemOpen
  \bibfield  {author} {\bibinfo {author} {\bibfnamefont {F.}~\bibnamefont
  {Sauter}},\ }\href {\doibase 10.1007/BF01339461} {\bibfield  {journal}
  {\bibinfo  {journal} {Z. Phys.}\ }\textbf {\bibinfo {volume} {69}},\ \bibinfo
  {pages} {742} (\bibinfo {year} {1931})}\BibitemShut {NoStop}%
\bibitem [{\citenamefont {Heisenberg}\ and\ \citenamefont
  {Euler}(1936)}]{Heisenberg:1935qt}%
  \BibitemOpen
  \bibfield  {author} {\bibinfo {author} {\bibfnamefont {W.}~\bibnamefont
  {Heisenberg}}\ and\ \bibinfo {author} {\bibfnamefont {H.}~\bibnamefont
  {Euler}},\ }\href {\doibase 10.1007/BF01343663, 10.1007/978-3-642-70078-1_9}
  {\bibfield  {journal} {\bibinfo  {journal} {Z. Phys.}\ }\textbf {\bibinfo
  {volume} {98}},\ \bibinfo {pages} {714} (\bibinfo {year} {1936})},\ \Eprint
  {http://arxiv.org/abs/physics/0605038} {arXiv:physics/0605038 [physics]}
  \BibitemShut {NoStop}%
\bibitem [{\citenamefont {Schwinger}(1951)}]{Schwinger:1951nm}%
  \BibitemOpen
  \bibfield  {author} {\bibinfo {author} {\bibfnamefont {J.~S.}\ \bibnamefont
  {Schwinger}},\ }\href {\doibase 10.1103/PhysRev.82.664} {\bibfield  {journal}
  {\bibinfo  {journal} {Phys. Rev.}\ }\textbf {\bibinfo {volume} {82}},\
  \bibinfo {pages} {664} (\bibinfo {year} {1951})}\BibitemShut {NoStop}%
\bibitem [{\citenamefont {Dunne}(2004)}]{Dunne:2004nc}%
  \BibitemOpen
  \bibfield  {author} {\bibinfo {author} {\bibfnamefont {G.~V.}\ \bibnamefont
  {Dunne}},\ }in\ \href {\doibase 10.1142/9789812775344_0014} {\emph {\bibinfo
  {booktitle} {From fields to strings: Circumnavigating theoretical physics.
  Ian Kogan memorial collection (3 volume set)}}},\ \bibinfo {editor} {edited
  by\ \bibinfo {editor} {\bibfnamefont {M.}~\bibnamefont {Shifman}}, \bibinfo
  {editor} {\bibfnamefont {A.}~\bibnamefont {Vainshtein}}, \ and\ \bibinfo
  {editor} {\bibfnamefont {J.}~\bibnamefont {Wheater}}}\ (\bibinfo {year}
  {2004})\ pp.\ \bibinfo {pages} {445--522},\ \Eprint
  {http://arxiv.org/abs/hep-th/0406216} {arXiv:hep-th/0406216 [hep-th]}
  \BibitemShut {NoStop}%
\bibitem [{\citenamefont {Gelis}\ and\ \citenamefont
  {Tanji}(2016)}]{Gelis:2015kya}%
  \BibitemOpen
  \bibfield  {author} {\bibinfo {author} {\bibfnamefont {F.}~\bibnamefont
  {Gelis}}\ and\ \bibinfo {author} {\bibfnamefont {N.}~\bibnamefont {Tanji}},\
  }\href {\doibase 10.1016/j.ppnp.2015.11.001} {\bibfield  {journal} {\bibinfo
  {journal} {Prog. Part. Nucl. Phys.}\ }\textbf {\bibinfo {volume} {87}},\
  \bibinfo {pages} {1} (\bibinfo {year} {2016})},\ \Eprint
  {http://arxiv.org/abs/1510.05451} {arXiv:1510.05451 [hep-ph]} \BibitemShut
  {NoStop}%
\bibitem [{\citenamefont {Parikh}\ and\ \citenamefont
  {Wilczek}(2000)}]{Parikh:1999mf}%
  \BibitemOpen
  \bibfield  {author} {\bibinfo {author} {\bibfnamefont {M.~K.}\ \bibnamefont
  {Parikh}}\ and\ \bibinfo {author} {\bibfnamefont {F.}~\bibnamefont
  {Wilczek}},\ }\href {\doibase 10.1103/PhysRevLett.85.5042} {\bibfield
  {journal} {\bibinfo  {journal} {Phys. Rev. Lett.}\ }\textbf {\bibinfo
  {volume} {85}},\ \bibinfo {pages} {5042} (\bibinfo {year} {2000})},\ \Eprint
  {http://arxiv.org/abs/hep-th/9907001} {arXiv:hep-th/9907001 [hep-th]}
  \BibitemShut {NoStop}%
\bibitem [{\citenamefont {Landau}(1932)}]{landau1932theorie}%
  \BibitemOpen
  \bibfield  {author} {\bibinfo {author} {\bibfnamefont {L.~D.}\ \bibnamefont
  {Landau}},\ }\href@noop {} {\bibfield  {journal} {\bibinfo  {journal} {Z.
  Sowjetunion}\ }\textbf {\bibinfo {volume} {2}},\ \bibinfo {pages} {46}
  (\bibinfo {year} {1932})}\BibitemShut {NoStop}%
\bibitem [{\citenamefont {Zener}\ and\ \citenamefont
  {Fowler}(1932)}]{zener1932non}%
  \BibitemOpen
  \bibfield  {author} {\bibinfo {author} {\bibfnamefont {C.}~\bibnamefont
  {Zener}}\ and\ \bibinfo {author} {\bibfnamefont {R.~H.}\ \bibnamefont
  {Fowler}},\ }\href {\doibase 10.1098/rspa.1932.0165} {\bibfield  {journal}
  {\bibinfo  {journal} {Proceedings of the Royal Society of London. Series A,
  Containing Papers of a Mathematical and Physical Character}\ }\textbf
  {\bibinfo {volume} {137}},\ \bibinfo {pages} {696} (\bibinfo {year}
  {1932})}\BibitemShut {NoStop}%
\bibitem [{\citenamefont {Dykhne}(1962)}]{dykhne1962adiabatic}%
  \BibitemOpen
  \bibfield  {author} {\bibinfo {author} {\bibfnamefont {A.}~\bibnamefont
  {Dykhne}},\ }\href@noop {} {\bibfield  {journal} {\bibinfo  {journal} {Sov.
  Phys. JETP}\ }\textbf {\bibinfo {volume} {14}},\ \bibinfo {pages} {1}
  (\bibinfo {year} {1962})}\BibitemShut {NoStop}%
\bibitem [{\citenamefont {Davis}\ and\ \citenamefont
  {Pechukas}(1976)}]{davis1976nonadiabatic}%
  \BibitemOpen
  \bibfield  {author} {\bibinfo {author} {\bibfnamefont {J.~P.}\ \bibnamefont
  {Davis}}\ and\ \bibinfo {author} {\bibfnamefont {P.}~\bibnamefont
  {Pechukas}},\ }\href {\doibase 10.1063/1.432648} {\bibfield  {journal}
  {\bibinfo  {journal} {The Journal of Chemical Physics}\ }\textbf {\bibinfo
  {volume} {64}},\ \bibinfo {pages} {3129} (\bibinfo {year}
  {1976})}\BibitemShut {NoStop}%
\bibitem [{\citenamefont {Suominen}\ \emph {et~al.}(1991)\citenamefont
  {Suominen}, \citenamefont {Garraway},\ and\ \citenamefont
  {Stenholm}}]{suominen1991adiabatic}%
  \BibitemOpen
  \bibfield  {author} {\bibinfo {author} {\bibfnamefont {K.-A.}\ \bibnamefont
  {Suominen}}, \bibinfo {author} {\bibfnamefont {B.}~\bibnamefont {Garraway}},
  \ and\ \bibinfo {author} {\bibfnamefont {S.}~\bibnamefont {Stenholm}},\
  }\href {\doibase https://doi.org/10.1016/0030-4018(91)90456-N} {\bibfield
  {journal} {\bibinfo  {journal} {Optics Communications}\ }\textbf {\bibinfo
  {volume} {82}},\ \bibinfo {pages} {260 } (\bibinfo {year}
  {1991})}\BibitemShut {NoStop}%
\bibitem [{\citenamefont {Vitanov}\ and\ \citenamefont
  {Suominen}(1999)}]{vitanov1999nonlinear}%
  \BibitemOpen
  \bibfield  {author} {\bibinfo {author} {\bibfnamefont {N.~V.}\ \bibnamefont
  {Vitanov}}\ and\ \bibinfo {author} {\bibfnamefont {K.-A.}\ \bibnamefont
  {Suominen}},\ }\href {\doibase 10.1103/PhysRevA.59.4580} {\bibfield
  {journal} {\bibinfo  {journal} {Phys. Rev. A}\ }\textbf {\bibinfo {volume}
  {59}},\ \bibinfo {pages} {4580} (\bibinfo {year} {1999})}\BibitemShut
  {NoStop}%
\bibitem [{\citenamefont {Wilkinson}\ and\ \citenamefont
  {Morgan}(2000)}]{wilkinson2000nonadiabatic}%
  \BibitemOpen
  \bibfield  {author} {\bibinfo {author} {\bibfnamefont {M.}~\bibnamefont
  {Wilkinson}}\ and\ \bibinfo {author} {\bibfnamefont {M.~A.}\ \bibnamefont
  {Morgan}},\ }\href {\doibase 10.1103/PhysRevA.61.062104} {\bibfield
  {journal} {\bibinfo  {journal} {Phys. Rev. A}\ }\textbf {\bibinfo {volume}
  {61}},\ \bibinfo {pages} {062104} (\bibinfo {year} {2000})}\BibitemShut
  {NoStop}%
\bibitem [{\citenamefont {Laine}\ and\ \citenamefont
  {Stenholm}(1996)}]{laine1996adiabatic}%
  \BibitemOpen
  \bibfield  {author} {\bibinfo {author} {\bibfnamefont {T.~A.}\ \bibnamefont
  {Laine}}\ and\ \bibinfo {author} {\bibfnamefont {S.}~\bibnamefont
  {Stenholm}},\ }\href {\doibase 10.1103/PhysRevA.53.2501} {\bibfield
  {journal} {\bibinfo  {journal} {Phys. Rev. A}\ }\textbf {\bibinfo {volume}
  {53}},\ \bibinfo {pages} {2501} (\bibinfo {year} {1996})}\BibitemShut
  {NoStop}%
\bibitem [{\citenamefont {Vitanov}\ and\ \citenamefont
  {Stenholm}(1996)}]{vitanov1996non}%
  \BibitemOpen
  \bibfield  {author} {\bibinfo {author} {\bibfnamefont {N.}~\bibnamefont
  {Vitanov}}\ and\ \bibinfo {author} {\bibfnamefont {S.}~\bibnamefont
  {Stenholm}},\ }\href {\doibase https://doi.org/10.1016/0030-4018(96)00216-7}
  {\bibfield  {journal} {\bibinfo  {journal} {Optics Communications}\ }\textbf
  {\bibinfo {volume} {127}},\ \bibinfo {pages} {215 } (\bibinfo {year}
  {1996})}\BibitemShut {NoStop}%
\bibitem [{\citenamefont {Witten}(2010)}]{Witten:2010zr}%
  \BibitemOpen
  \bibfield  {author} {\bibinfo {author} {\bibfnamefont {E.}~\bibnamefont
  {Witten}},\ }\href@noop {} {\  (\bibinfo {year} {2010})},\ \Eprint
  {http://arxiv.org/abs/1009.6032} {arXiv:1009.6032 [hep-th]} \BibitemShut
  {NoStop}%
\bibitem [{\citenamefont {Witten}(2011)}]{witten2011analytic}%
  \BibitemOpen
  \bibfield  {author} {\bibinfo {author} {\bibfnamefont {E.}~\bibnamefont
  {Witten}},\ }\bibfield  {booktitle} {\emph {\bibinfo {booktitle}
  {{Chern-Simons gauge theory: 20 years after. Proceedings, Workshop, Bonn,
  Germany, August 3-7, 2009}}},\ }\href@noop {} {\bibfield  {journal} {\bibinfo
   {journal} {AMS/IP Stud. Adv. Math.}\ }\textbf {\bibinfo {volume} {50}},\
  \bibinfo {pages} {347} (\bibinfo {year} {2011})},\ \Eprint
  {http://arxiv.org/abs/1001.2933} {arXiv:1001.2933 [hep-th]} \BibitemShut
  {NoStop}%
\bibitem [{\citenamefont {Cristoforetti}\ \emph {et~al.}(2012)\citenamefont
  {Cristoforetti}, \citenamefont {Di~Renzo},\ and\ \citenamefont
  {Scorzato}}]{Cristoforetti:2012su}%
  \BibitemOpen
  \bibfield  {author} {\bibinfo {author} {\bibfnamefont {M.}~\bibnamefont
  {Cristoforetti}}, \bibinfo {author} {\bibfnamefont {F.}~\bibnamefont
  {Di~Renzo}}, \ and\ \bibinfo {author} {\bibfnamefont {L.}~\bibnamefont
  {Scorzato}} (\bibinfo {collaboration} {AuroraScience}),\ }\href {\doibase
  10.1103/PhysRevD.86.074506} {\bibfield  {journal} {\bibinfo  {journal} {Phys.
  Rev.}\ }\textbf {\bibinfo {volume} {D86}},\ \bibinfo {pages} {074506}
  (\bibinfo {year} {2012})},\ \Eprint {http://arxiv.org/abs/1205.3996}
  {arXiv:1205.3996 [hep-lat]} \BibitemShut {NoStop}%
\bibitem [{\citenamefont {Cristoforetti}\ \emph {et~al.}(2013)\citenamefont
  {Cristoforetti}, \citenamefont {Di~Renzo}, \citenamefont {Mukherjee},\ and\
  \citenamefont {Scorzato}}]{Cristoforetti:2013wha}%
  \BibitemOpen
  \bibfield  {author} {\bibinfo {author} {\bibfnamefont {M.}~\bibnamefont
  {Cristoforetti}}, \bibinfo {author} {\bibfnamefont {F.}~\bibnamefont
  {Di~Renzo}}, \bibinfo {author} {\bibfnamefont {A.}~\bibnamefont {Mukherjee}},
  \ and\ \bibinfo {author} {\bibfnamefont {L.}~\bibnamefont {Scorzato}},\
  }\href {\doibase 10.1103/PhysRevD.88.051501} {\bibfield  {journal} {\bibinfo
  {journal} {Phys. Rev.}\ }\textbf {\bibinfo {volume} {D88}},\ \bibinfo {pages}
  {051501} (\bibinfo {year} {2013})},\ \Eprint {http://arxiv.org/abs/1303.7204}
  {arXiv:1303.7204 [hep-lat]} \BibitemShut {NoStop}%
\bibitem [{\citenamefont {Fujii}\ \emph {et~al.}(2013)\citenamefont {Fujii},
  \citenamefont {Honda}, \citenamefont {Kato}, \citenamefont {Kikukawa},
  \citenamefont {Komatsu},\ and\ \citenamefont {Sano}}]{Fujii:2013sra}%
  \BibitemOpen
  \bibfield  {author} {\bibinfo {author} {\bibfnamefont {H.}~\bibnamefont
  {Fujii}}, \bibinfo {author} {\bibfnamefont {D.}~\bibnamefont {Honda}},
  \bibinfo {author} {\bibfnamefont {M.}~\bibnamefont {Kato}}, \bibinfo {author}
  {\bibfnamefont {Y.}~\bibnamefont {Kikukawa}}, \bibinfo {author}
  {\bibfnamefont {S.}~\bibnamefont {Komatsu}}, \ and\ \bibinfo {author}
  {\bibfnamefont {T.}~\bibnamefont {Sano}},\ }\href {\doibase
  10.1007/JHEP10(2013)147} {\bibfield  {journal} {\bibinfo  {journal} {JHEP}\
  }\textbf {\bibinfo {volume} {10}},\ \bibinfo {pages} {147} (\bibinfo {year}
  {2013})},\ \Eprint {http://arxiv.org/abs/1309.4371} {arXiv:1309.4371
  [hep-lat]} \BibitemShut {NoStop}%
\bibitem [{\citenamefont {Tanizaki}\ \emph {et~al.}(2016)\citenamefont
  {Tanizaki}, \citenamefont {Hidaka},\ and\ \citenamefont
  {Hayata}}]{tanizaki2016lefschetz}%
  \BibitemOpen
  \bibfield  {author} {\bibinfo {author} {\bibfnamefont {Y.}~\bibnamefont
  {Tanizaki}}, \bibinfo {author} {\bibfnamefont {Y.}~\bibnamefont {Hidaka}}, \
  and\ \bibinfo {author} {\bibfnamefont {T.}~\bibnamefont {Hayata}},\ }\href
  {\doibase 10.1088/1367-2630/18/3/033002} {\bibfield  {journal} {\bibinfo
  {journal} {New J. Phys.}\ }\textbf {\bibinfo {volume} {18}},\ \bibinfo
  {pages} {033002} (\bibinfo {year} {2016})},\ \Eprint
  {http://arxiv.org/abs/1509.07146} {arXiv:1509.07146 [hep-th]} \BibitemShut
  {NoStop}%
\bibitem [{\citenamefont {Tanizaki}\ and\ \citenamefont
  {Koike}(2014)}]{tanizaki2014real}%
  \BibitemOpen
  \bibfield  {author} {\bibinfo {author} {\bibfnamefont {Y.}~\bibnamefont
  {Tanizaki}}\ and\ \bibinfo {author} {\bibfnamefont {T.}~\bibnamefont
  {Koike}},\ }\href {\doibase 10.1016/j.aop.2014.09.003} {\bibfield  {journal}
  {\bibinfo  {journal} {Annals Phys.}\ }\textbf {\bibinfo {volume} {351}},\
  \bibinfo {pages} {250} (\bibinfo {year} {2014})},\ \Eprint
  {http://arxiv.org/abs/1406.2386} {arXiv:1406.2386 [math-ph]} \BibitemShut
  {NoStop}%
\bibitem [{\citenamefont {Cherman}\ and\ \citenamefont
  {Unsal}(2014)}]{Cherman:2014sba}%
  \BibitemOpen
  \bibfield  {author} {\bibinfo {author} {\bibfnamefont {A.}~\bibnamefont
  {Cherman}}\ and\ \bibinfo {author} {\bibfnamefont {M.}~\bibnamefont
  {Unsal}},\ }\href@noop {} {\  (\bibinfo {year} {2014})},\ \Eprint
  {http://arxiv.org/abs/1408.0012} {arXiv:1408.0012 [hep-th]} \BibitemShut
  {NoStop}%
\bibitem [{\citenamefont {Andreassen}\ \emph {et~al.}(2017)\citenamefont
  {Andreassen}, \citenamefont {Farhi}, \citenamefont {Frost},\ and\
  \citenamefont {Schwartz}}]{andreassen2017precision}%
  \BibitemOpen
  \bibfield  {author} {\bibinfo {author} {\bibfnamefont {A.}~\bibnamefont
  {Andreassen}}, \bibinfo {author} {\bibfnamefont {D.}~\bibnamefont {Farhi}},
  \bibinfo {author} {\bibfnamefont {W.}~\bibnamefont {Frost}}, \ and\ \bibinfo
  {author} {\bibfnamefont {M.~D.}\ \bibnamefont {Schwartz}},\ }\href {\doibase
  10.1103/PhysRevD.95.085011} {\bibfield  {journal} {\bibinfo  {journal} {Phys.
  Rev.}\ }\textbf {\bibinfo {volume} {D95}},\ \bibinfo {pages} {085011}
  (\bibinfo {year} {2017})},\ \Eprint {http://arxiv.org/abs/1604.06090}
  {arXiv:1604.06090 [hep-th]} \BibitemShut {NoStop}%
\bibitem [{\citenamefont {Behtash}\ \emph {et~al.}(2017)\citenamefont
  {Behtash}, \citenamefont {Dunne}, \citenamefont {Schaefer}, \citenamefont
  {Sulejmanpasic},\ and\ \citenamefont {Unsal}}]{Behtash:2015loa}%
  \BibitemOpen
  \bibfield  {author} {\bibinfo {author} {\bibfnamefont {A.}~\bibnamefont
  {Behtash}}, \bibinfo {author} {\bibfnamefont {G.~V.}\ \bibnamefont {Dunne}},
  \bibinfo {author} {\bibfnamefont {T.}~\bibnamefont {Schaefer}}, \bibinfo
  {author} {\bibfnamefont {T.}~\bibnamefont {Sulejmanpasic}}, \ and\ \bibinfo
  {author} {\bibfnamefont {M.}~\bibnamefont {Unsal}},\ }\href {\doibase
  10.4310/AMSA.2017.v2.n1.a3} {\bibfield  {journal} {\bibinfo  {journal}
  {Annals of Mathematical Sciences and Applications}\ }\textbf {\bibinfo
  {volume} {Volume 2}},\ \bibinfo {pages} {No. 1} (\bibinfo {year} {2017})},\
  \Eprint {http://arxiv.org/abs/1510.03435} {arXiv:1510.03435 [hep-th]}
  \BibitemShut {NoStop}%
\bibitem [{\citenamefont {Gould}\ and\ \citenamefont
  {Rajantie}(2017)}]{Gould:2017fve}%
  \BibitemOpen
  \bibfield  {author} {\bibinfo {author} {\bibfnamefont {O.}~\bibnamefont
  {Gould}}\ and\ \bibinfo {author} {\bibfnamefont {A.}~\bibnamefont
  {Rajantie}},\ }\href {\doibase 10.1103/PhysRevD.96.076002} {\bibfield
  {journal} {\bibinfo  {journal} {Phys. Rev.}\ }\textbf {\bibinfo {volume}
  {D96}},\ \bibinfo {pages} {076002} (\bibinfo {year} {2017})},\ \Eprint
  {http://arxiv.org/abs/1704.04801} {arXiv:1704.04801 [hep-th]} \BibitemShut
  {NoStop}%
\bibitem [{\citenamefont {Draper}(2018)}]{Draper:2018lyw}%
  \BibitemOpen
  \bibfield  {author} {\bibinfo {author} {\bibfnamefont {P.}~\bibnamefont
  {Draper}},\ }\href {\doibase 10.1103/PhysRevD.98.125014} {\bibfield
  {journal} {\bibinfo  {journal} {Phys. Rev.}\ }\textbf {\bibinfo {volume}
  {D98}},\ \bibinfo {pages} {125014} (\bibinfo {year} {2018})},\ \Eprint
  {http://arxiv.org/abs/1809.10768} {arXiv:1809.10768 [hep-th]} \BibitemShut
  {NoStop}%
\bibitem [{\citenamefont {Gould}\ \emph {et~al.}(2019)\citenamefont {Gould},
  \citenamefont {Mangles}, \citenamefont {Rajantie}, \citenamefont {Rose},\
  and\ \citenamefont {Xie}}]{Gould:2018efv}%
  \BibitemOpen
  \bibfield  {author} {\bibinfo {author} {\bibfnamefont {O.}~\bibnamefont
  {Gould}}, \bibinfo {author} {\bibfnamefont {S.}~\bibnamefont {Mangles}},
  \bibinfo {author} {\bibfnamefont {A.}~\bibnamefont {Rajantie}}, \bibinfo
  {author} {\bibfnamefont {S.}~\bibnamefont {Rose}}, \ and\ \bibinfo {author}
  {\bibfnamefont {C.}~\bibnamefont {Xie}},\ }\href {\doibase
  10.1103/PhysRevA.99.052120} {\bibfield  {journal} {\bibinfo  {journal} {Phys.
  Rev.}\ }\textbf {\bibinfo {volume} {A99}},\ \bibinfo {pages} {052120}
  (\bibinfo {year} {2019})},\ \Eprint {http://arxiv.org/abs/1812.04089}
  {arXiv:1812.04089 [hep-ph]} \BibitemShut {NoStop}%
\bibitem [{\citenamefont {Gould}\ \emph {et~al.}(2018)\citenamefont {Gould},
  \citenamefont {Rajantie},\ and\ \citenamefont {Xie}}]{Gould:2018ovk}%
  \BibitemOpen
  \bibfield  {author} {\bibinfo {author} {\bibfnamefont {O.}~\bibnamefont
  {Gould}}, \bibinfo {author} {\bibfnamefont {A.}~\bibnamefont {Rajantie}}, \
  and\ \bibinfo {author} {\bibfnamefont {C.}~\bibnamefont {Xie}},\ }\href
  {\doibase 10.1103/PhysRevD.98.056022} {\bibfield  {journal} {\bibinfo
  {journal} {Phys. Rev.}\ }\textbf {\bibinfo {volume} {D98}},\ \bibinfo {pages}
  {056022} (\bibinfo {year} {2018})},\ \Eprint
  {http://arxiv.org/abs/1806.02665} {arXiv:1806.02665 [hep-th]} \BibitemShut
  {NoStop}%
\bibitem [{\citenamefont {Schutzhold}\ \emph {et~al.}(2008)\citenamefont
  {Schutzhold}, \citenamefont {Gies},\ and\ \citenamefont
  {Dunne}}]{Schutzhold:2008pz}%
  \BibitemOpen
  \bibfield  {author} {\bibinfo {author} {\bibfnamefont {R.}~\bibnamefont
  {Schutzhold}}, \bibinfo {author} {\bibfnamefont {H.}~\bibnamefont {Gies}}, \
  and\ \bibinfo {author} {\bibfnamefont {G.}~\bibnamefont {Dunne}},\ }\href
  {\doibase 10.1103/PhysRevLett.101.130404} {\bibfield  {journal} {\bibinfo
  {journal} {Phys. Rev. Lett.}\ }\textbf {\bibinfo {volume} {101}},\ \bibinfo
  {pages} {130404} (\bibinfo {year} {2008})},\ \Eprint
  {http://arxiv.org/abs/0807.0754} {arXiv:0807.0754 [hep-th]} \BibitemShut
  {NoStop}%
\bibitem [{\citenamefont {Dunne}\ \emph {et~al.}(2009)\citenamefont {Dunne},
  \citenamefont {Gies},\ and\ \citenamefont {Schutzhold}}]{Dunne:2009gi}%
  \BibitemOpen
  \bibfield  {author} {\bibinfo {author} {\bibfnamefont {G.~V.}\ \bibnamefont
  {Dunne}}, \bibinfo {author} {\bibfnamefont {H.}~\bibnamefont {Gies}}, \ and\
  \bibinfo {author} {\bibfnamefont {R.}~\bibnamefont {Schutzhold}},\ }\href
  {\doibase 10.1103/PhysRevD.80.111301} {\bibfield  {journal} {\bibinfo
  {journal} {Phys. Rev.}\ }\textbf {\bibinfo {volume} {D80}},\ \bibinfo {pages}
  {111301} (\bibinfo {year} {2009})},\ \Eprint {http://arxiv.org/abs/0908.0948}
  {arXiv:0908.0948 [hep-ph]} \BibitemShut {NoStop}%
\bibitem [{\citenamefont {Orthaber}\ \emph {et~al.}(2011)\citenamefont
  {Orthaber}, \citenamefont {Hebenstreit},\ and\ \citenamefont
  {Alkofer}}]{Orthaber:2011cm}%
  \BibitemOpen
  \bibfield  {author} {\bibinfo {author} {\bibfnamefont {M.}~\bibnamefont
  {Orthaber}}, \bibinfo {author} {\bibfnamefont {F.}~\bibnamefont
  {Hebenstreit}}, \ and\ \bibinfo {author} {\bibfnamefont {R.}~\bibnamefont
  {Alkofer}},\ }\href {\doibase 10.1016/j.physletb.2011.02.053} {\bibfield
  {journal} {\bibinfo  {journal} {Phys. Lett.}\ }\textbf {\bibinfo {volume}
  {B698}},\ \bibinfo {pages} {80} (\bibinfo {year} {2011})},\ \Eprint
  {http://arxiv.org/abs/1102.2182} {arXiv:1102.2182 [hep-ph]} \BibitemShut
  {NoStop}%
\bibitem [{\citenamefont {Fey}\ and\ \citenamefont
  {Schutzhold}(2012)}]{Fey:2011if}%
  \BibitemOpen
  \bibfield  {author} {\bibinfo {author} {\bibfnamefont {C.}~\bibnamefont
  {Fey}}\ and\ \bibinfo {author} {\bibfnamefont {R.}~\bibnamefont
  {Schutzhold}},\ }\href {\doibase 10.1103/PhysRevD.85.025004} {\bibfield
  {journal} {\bibinfo  {journal} {Phys. Rev.}\ }\textbf {\bibinfo {volume}
  {D85}},\ \bibinfo {pages} {025004} (\bibinfo {year} {2012})},\ \Eprint
  {http://arxiv.org/abs/1110.5499} {arXiv:1110.5499 [hep-th]} \BibitemShut
  {NoStop}%
\bibitem [{\citenamefont {Li}\ \emph {et~al.}(2014)\citenamefont {Li},
  \citenamefont {Lu}, \citenamefont {Xie}, \citenamefont {Fu}, \citenamefont
  {Liu},\ and\ \citenamefont {Shen}}]{Li:2014psw}%
  \BibitemOpen
  \bibfield  {author} {\bibinfo {author} {\bibfnamefont {Z.~L.}\ \bibnamefont
  {Li}}, \bibinfo {author} {\bibfnamefont {D.}~\bibnamefont {Lu}}, \bibinfo
  {author} {\bibfnamefont {B.~S.}\ \bibnamefont {Xie}}, \bibinfo {author}
  {\bibfnamefont {L.~B.}\ \bibnamefont {Fu}}, \bibinfo {author} {\bibfnamefont
  {J.}~\bibnamefont {Liu}}, \ and\ \bibinfo {author} {\bibfnamefont {B.~F.}\
  \bibnamefont {Shen}},\ }\href {\doibase 10.1103/PhysRevD.89.093011}
  {\bibfield  {journal} {\bibinfo  {journal} {Phys. Rev.}\ }\textbf {\bibinfo
  {volume} {D89}},\ \bibinfo {pages} {093011} (\bibinfo {year}
  {2014})}\BibitemShut {NoStop}%
\bibitem [{\citenamefont {Copinger}\ and\ \citenamefont
  {Fukushima}(2016)}]{Copinger:2016llk}%
  \BibitemOpen
  \bibfield  {author} {\bibinfo {author} {\bibfnamefont {P.}~\bibnamefont
  {Copinger}}\ and\ \bibinfo {author} {\bibfnamefont {K.}~\bibnamefont
  {Fukushima}},\ }\href {\doibase 10.1103/PhysRevLett.117.081603,
  10.1103/PhysRevLett.118.099903} {\bibfield  {journal} {\bibinfo  {journal}
  {Phys. Rev. Lett.}\ }\textbf {\bibinfo {volume} {117}},\ \bibinfo {pages}
  {081603} (\bibinfo {year} {2016})},\ \bibinfo {note} {[Erratum: Phys. Rev.
  Lett.118,no.9,099903(2017)]},\ \Eprint {http://arxiv.org/abs/1605.05957}
  {arXiv:1605.05957 [hep-th]} \BibitemShut {NoStop}%
\bibitem [{\citenamefont {Schneider}\ and\ \citenamefont
  {Schutzhold}(2016)}]{Schneider:2016vrl}%
  \BibitemOpen
  \bibfield  {author} {\bibinfo {author} {\bibfnamefont {C.}~\bibnamefont
  {Schneider}}\ and\ \bibinfo {author} {\bibfnamefont {R.}~\bibnamefont
  {Schutzhold}},\ }\href {\doibase 10.1103/PhysRevD.94.085015} {\bibfield
  {journal} {\bibinfo  {journal} {Phys. Rev.}\ }\textbf {\bibinfo {volume}
  {D94}},\ \bibinfo {pages} {085015} (\bibinfo {year} {2016})},\ \Eprint
  {http://arxiv.org/abs/1603.00864} {arXiv:1603.00864 [hep-th]} \BibitemShut
  {NoStop}%
\bibitem [{\citenamefont {Torgrimsson}\ \emph {et~al.}(2017)\citenamefont
  {Torgrimsson}, \citenamefont {Schneider}, \citenamefont {Oertel},\ and\
  \citenamefont {Schutzhold}}]{Torgrimsson:2017pzs}%
  \BibitemOpen
  \bibfield  {author} {\bibinfo {author} {\bibfnamefont {G.}~\bibnamefont
  {Torgrimsson}}, \bibinfo {author} {\bibfnamefont {C.}~\bibnamefont
  {Schneider}}, \bibinfo {author} {\bibfnamefont {J.}~\bibnamefont {Oertel}}, \
  and\ \bibinfo {author} {\bibfnamefont {R.}~\bibnamefont {Schutzhold}},\
  }\href {\doibase 10.1007/JHEP06(2017)043} {\bibfield  {journal} {\bibinfo
  {journal} {JHEP}\ }\textbf {\bibinfo {volume} {06}},\ \bibinfo {pages} {043}
  (\bibinfo {year} {2017})},\ \Eprint {http://arxiv.org/abs/1703.09203}
  {arXiv:1703.09203 [hep-th]} \BibitemShut {NoStop}%
\bibitem [{\citenamefont {Torgrimsson}\ \emph {et~al.}(2018)\citenamefont
  {Torgrimsson}, \citenamefont {Schneider},\ and\ \citenamefont
  {Schutzhold}}]{Torgrimsson:2017cyb}%
  \BibitemOpen
  \bibfield  {author} {\bibinfo {author} {\bibfnamefont {G.}~\bibnamefont
  {Torgrimsson}}, \bibinfo {author} {\bibfnamefont {C.}~\bibnamefont
  {Schneider}}, \ and\ \bibinfo {author} {\bibfnamefont {R.}~\bibnamefont
  {Schutzhold}},\ }\href {\doibase 10.1103/PhysRevD.97.096004} {\bibfield
  {journal} {\bibinfo  {journal} {Phys. Rev.}\ }\textbf {\bibinfo {volume}
  {D97}},\ \bibinfo {pages} {096004} (\bibinfo {year} {2018})},\ \Eprint
  {http://arxiv.org/abs/1712.08613} {arXiv:1712.08613 [hep-ph]} \BibitemShut
  {NoStop}%
\bibitem [{\citenamefont {Berry}(1990)}]{berry1990histories}%
  \BibitemOpen
  \bibfield  {author} {\bibinfo {author} {\bibfnamefont {M.~V.}\ \bibnamefont
  {Berry}},\ }\href {\doibase 10.1098/rspa.1990.0051} {\bibfield  {journal}
  {\bibinfo  {journal} {Proceedings of the Royal Society of London. A.
  Mathematical and Physical Sciences}\ }\textbf {\bibinfo {volume} {429}},\
  \bibinfo {pages} {61} (\bibinfo {year} {1990})}\BibitemShut {NoStop}%
\bibitem [{\citenamefont {Cohen}\ and\ \citenamefont
  {McGady}(2008)}]{Cohen:2008wz}%
  \BibitemOpen
  \bibfield  {author} {\bibinfo {author} {\bibfnamefont {T.~D.}\ \bibnamefont
  {Cohen}}\ and\ \bibinfo {author} {\bibfnamefont {D.~A.}\ \bibnamefont
  {McGady}},\ }\href {\doibase 10.1103/PhysRevD.78.036008} {\bibfield
  {journal} {\bibinfo  {journal} {Phys. Rev.}\ }\textbf {\bibinfo {volume}
  {D78}},\ \bibinfo {pages} {036008} (\bibinfo {year} {2008})},\ \Eprint
  {http://arxiv.org/abs/0807.1117} {arXiv:0807.1117 [hep-ph]} \BibitemShut
  {NoStop}%
\bibitem [{\citenamefont {Rosen}\ and\ \citenamefont
  {Zener}(1932)}]{rosen1932double}%
  \BibitemOpen
  \bibfield  {author} {\bibinfo {author} {\bibfnamefont {N.}~\bibnamefont
  {Rosen}}\ and\ \bibinfo {author} {\bibfnamefont {C.}~\bibnamefont {Zener}},\
  }\href {\doibase 10.1103/PhysRev.40.502} {\bibfield  {journal} {\bibinfo
  {journal} {Phys. Rev.}\ }\textbf {\bibinfo {volume} {40}},\ \bibinfo {pages}
  {502} (\bibinfo {year} {1932})}\BibitemShut {NoStop}%
\bibitem [{\citenamefont {Fukushima}\ \emph {et~al.}(2009)\citenamefont
  {Fukushima}, \citenamefont {Gelis},\ and\ \citenamefont
  {Lappi}}]{Fukushima:2009er}%
  \BibitemOpen
  \bibfield  {author} {\bibinfo {author} {\bibfnamefont {K.}~\bibnamefont
  {Fukushima}}, \bibinfo {author} {\bibfnamefont {F.}~\bibnamefont {Gelis}}, \
  and\ \bibinfo {author} {\bibfnamefont {T.}~\bibnamefont {Lappi}},\ }\href
  {\doibase 10.1016/j.nuclphysa.2009.09.062} {\bibfield  {journal} {\bibinfo
  {journal} {Nucl. Phys.}\ }\textbf {\bibinfo {volume} {A831}},\ \bibinfo
  {pages} {184} (\bibinfo {year} {2009})},\ \Eprint
  {http://arxiv.org/abs/0907.4793} {arXiv:0907.4793 [hep-ph]} \BibitemShut
  {NoStop}%
\end{thebibliography}%
\bibliographystyle{apsrev4-1}

\end{document}